# Intrinsically chiral ferronematic liquid crystals

D. Pociecha,[a] R. Walker,[b] E. Cruickshank,[b] J. Szydlowska,[a] P. Rybak,[a] A. Makal,[a] J. Matraszek,[a] J. M. Wolska,[a] J.M.D. Storey,[b] C.T. Imrie [b] and E. Gorecka*[a]

[a] Department of Chemistry, University of Warsaw, Zwirki i Wigury 101, 02-089 Warsaw, Poland
[b] Department of Chemistry, University of Aberdeen, Old Aberdeen AB24 3UE, U.K.

**Abstract:** Strongly dipolar mesogenic compounds with a chiral center located in a lateral alkyl chain were synthesized, and shown to form the ferroelectric nematic phase. The presence of molecular chirality induced a helical structure in both the N and $N_F$ phases, but with opposite helix sense in the two phases. The relaxation frequency of the polar fluctuations was found to be lower for the chiral $N_F$ phase than for its achiral, non-branched counterpart with the same lateral chain length.

Chirality is a dominant feature of many biologically active molecules that are capable of forming helical assemblies even in the liquid state. In fluidic liquid crystalline phases, chiral molecules twist with respect to each other and this is transmitted over much longer length scales giving helical phase structures. The sense of the helix is pre-determined by the sense of molecular chirality and molecular conformation [1]. Thus, in the chiral nematic (cholesteric) phase, a helical distribution of the director is obtained in the absence of any positional order. By contrast, to obtain a polar phase, in which there is long-range ordering of dipole moments, usually requires at least some degree of positional order. Until very recently, the combination of chirality and polar order in soft matter had been observed only in smectic [2] and columnar phases [3]. The recent discovery of a nematic phase with ferroelectric order [4] raises the fundamental question, how will intrinsic molecular chirality influence the structure of the ferroelectric nematic ($N_F$) phase? To date, the only reports of chiral $N_F$ phases have been for achiral nematogens doped with chiral additives [5]. Here we report, for the first time, the synthesis and characterization of intrinsically chiral ferroelectric nematogens, and compare their properties to those of structurally similar achiral materials. These compounds are based on the standard ferroelectric nematogen, RM-734 [4d], here referred to as **I-1**, and their structures and transitional properties are shown in Table 1. The chiral center has been introduced into a laterally attached alkyl chain at the third atom away from the mesogenic core. The single crystal x-ray diffraction data for compounds **I-3** and **I-4**, the only materials for which good quality crystals could be obtained, reveals that in the crystal lattice (monoclinic $P2_1/c$ space group) the lateral chains are in a fully extended conformation, and the electric dipoles of neighboring molecules are arranged in an antiparallel fashion (see SI).

The homologous series **I-m**, in which the lateral *n*-alkyl chain increases from methyl to hexyl was studied to guide the design of the target chiral ferroelectric nematogen, compound **I-4***. All the homologues show monotropic nematic and ferro-nematic phases, except for the methyl member, **I-1**, for which the nematic phase was enantiotropic (Table 1). It must be noted, however, that all these materials readily supercool to temperatures far below their melting points. Indeed, in some cases the monotropic $N_F$ phase was stable at room temperature for hours prior to crystallizing. As expected, the Iso-N and N-$N_F$ transition temperatures decrease as the lateral alkyl chain is extended, and the effect on the clearing temperature is more pronounced.

**Table 1.** Molecular structures and transition temperatures obtained from DSC scans (10 K/min). The associated enthalpy changes are given in brackets (J/g).

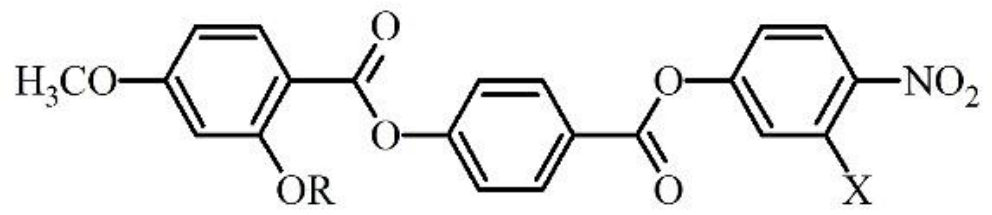

| | R | X | |
|---|---|---|---|
| I-1 | $CH_3$ | H | Cr 139.0(80.5) N 188.0 (1.50) Iso<br>Iso 187.0 (1.50) N 130.7 (1.50) $N_F$ |
| I-2 | $C_2H_5$ | H | Cr 159.0 (131.0) Iso<br>Iso 131.0 (1.90) N 106.0 (3.60) N $_F$ |
| I-3 | $C_3H_7$ | H | Cr 146.0(98.9) Iso<br>Iso 95.7 (3.20) N 84.6 (1.50) $N_F$ |
| I-4 | $C_4H_9$ | H | Cr 141.0(97.8) Iso<br>Iso 74.6 (0.63) N 64.8 (1.75) $N_F$ |
| I-5 | $C_5H_{11}$ | H | Cr 130.4 (108.2) Iso<br>Iso 60.0(1.05) N 52.7 (2.20) $N_F$ |
| I-6 | $C_6H_{13}$ | H | Cr 101.0 (73.8) Iso<br>Iso 50.1(0.76) N 43.6 (1.27) $N_F$ |
| I-4* | $CH_2CH(CH_3)$<br>$C_2H_5$ | H | Cr 140.0 (92.1) Iso<br>Iso 56.1 (-) N* 54.8 (3.48†) $N^*_F$ |
| II-6 | $C_6H_{13}$ | F | Cr 77.0 (75.5) Iso<br>Iso 47.8 (4.20) $N_F$ |
| II-4* | $CH_2CH(CH_3)$<br>$C_2H_5$ | F | Cr 111.0 (74.8) Iso<br>Iso 54.9 (5.20) $N^*_F$ |

† the peaks corresponding to the N-$N_F$ and Iso-N transitions are not well resolved.

As a result, the temperature range of the higher temperature nematic phase decreases with increasing lateral chain length (inset in Fig. 1). Both transition temperatures appear to be approaching limiting values on increasing chain length, and for $T_{NI}$ such behavior is normally attributed to the alkyl chain adopting conformations in which it lies along the principal molecular axis [6]. The overall decrease in both transition temperatures clearly correlates with a decreasing molecular shape anisotropy, as evidenced by the optical birefringence measurements (Fig. 1). Specifically, $\Delta n$ strongly decreases in both the N and $N_F$ phases as the lateral chain is increased (Fig. 1).

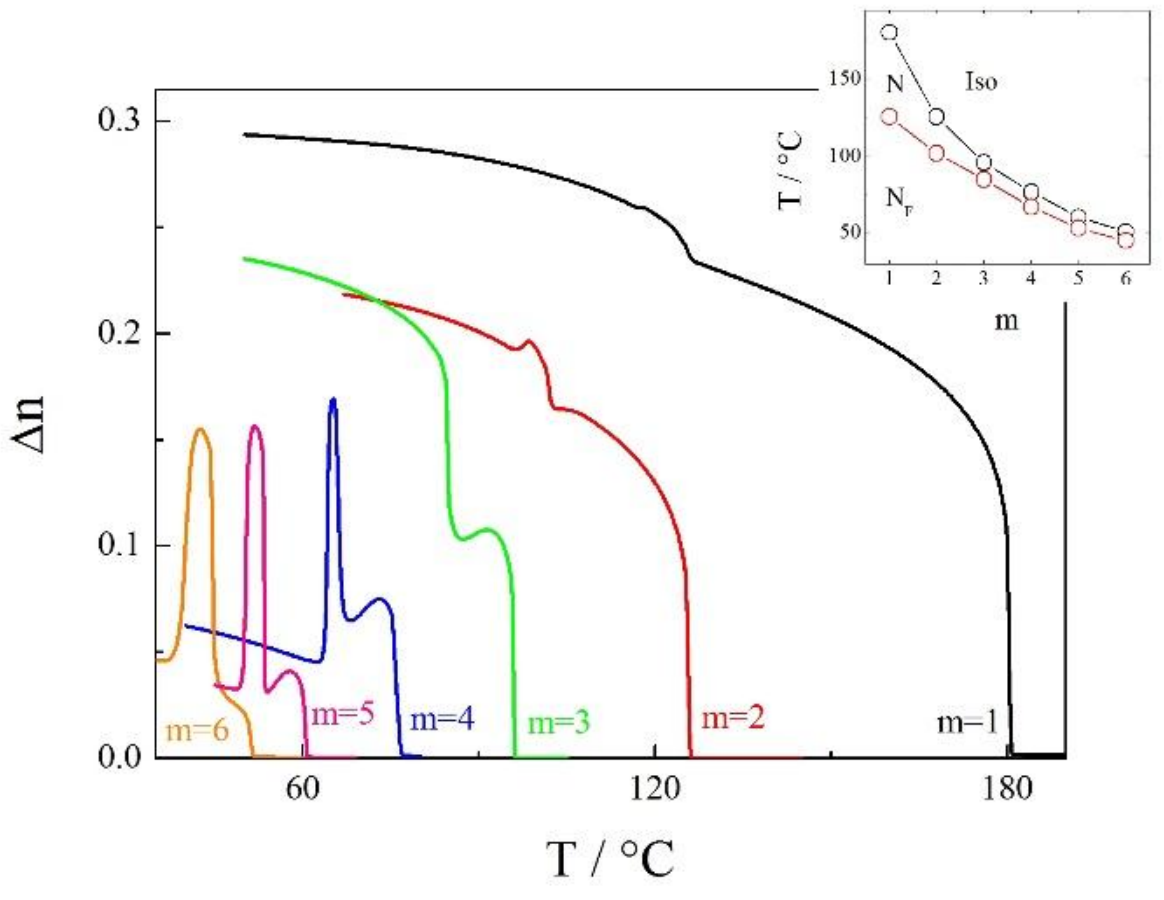

**Figure 1.** Birefringence measured with green light (532nm) as a function of temperature for homologues **I-m** with different lateral chain lengths, m; a decrease in structural anisotropy results in smaller values of the birefringence. At the N-N$_F$ phase transition, there is a step-wise increase of birefringence because of the higher order parameter. A decrease in birefringence observed in the N$_F$ phase a few degrees below the transition from the N phase, for m=4-6 (**I-4, I-5** and **I-6**) is due to the formation of twisted states induced by interactions with the antiparallel rubbed polymer layers at the cell surfaces.

In addition, as the temperature range of the N phase preceding the N$_F$ phase decreases, the N-N$_F$ transition becomes more strongly first order in character, and is accompanied by a larger jump in Δn. Hence, a stronger increase in the order parameter at this transition is detected (Fig. 1). A further decrease in the molecular anisotropy by introducing an F atom *ortho* to the terminal nitro group (compounds **II**) destabilized the N phase completely, and a direct Iso-N$_F$ phase transition was observed for compound **II-6**. Branching the lateral chain by introducing a chiral carbon atom (*cf* compounds **I-4** and **I-4***) decreases the nematic phase stabilities even further, both the N$^*$- N$^*_F$ and Iso-N$^*$ phase transition temperatures for the chiral compound **I-4*** are lower than for its unbranched counterpart **I-4** by 10 and 19 K, respectively. Introducing the additional F substituent *ortho* to the terminal nitro group in the mesogenic core (compound **II-4***) further destabilized the N* phase revealing a direct Iso-N$^*_F$ transition with a clearing temperature similar to the non-substituted chiral material **I-4***. It should be noted that compound **I-4*** was recently mentioned in ref. [5b], but its mesogenic properties were missed, the material was used only as a chiral dopant. Molecular chirality introduces a twist deformation of the director (*i.e.* the averaged local direction of the long molecular axes) field and both nematic phases acquire a helical structure. The helical pitch in the N$^*_F$ phase could be deduced from its optical texture. Thus, while fast cooling in a few micron thick cell treated for planar anchoring resulted in uniform orientation of helix perpendicular to the cell surface (Fig. S5), slow cooling of the sample resulted in a well-defined fingerprint texture, with a periodicity of ~2.5 μm and ~5 μm for compounds **I-4*** and **II-4***, respectively (Fig. 2 – inset, Fig. S5).

This suggests that for wavelengths in the visible region, the optical rotatory power (ORP) is in the Mauguin regime (for λ<np), and therefore, shows a linear dependence on the helical pitch length according to: $ORP=-\pi(\delta n)^2p/4\lambda^4$ [7].

For green light (532 nm) for compound **I-4***, the ORP was ~8 deg./μm in the N$^*_F$ phase and ~3 deg./ μm in the N* phase; assuming that the birefringence of the nematic phases are similar to that of the achiral counterpart **I-4**, and taking into account the difference in Δn between the nematic and polar nematic phases, the helical pitch calculated from the ORP values in both the N* and N$^*_F$ phases is ~ 2 μm.

For compound **II-4*** the value of ORP reaches 15 deg./ μm, this larger optical activity is consistent with the longer pitch in this material. Interestingly, the ORP have opposite signs in the N$^*_F$ and N$^*$ phases (Fig. 2), showing that the helices in these phases have opposite sense. This might result from the different interactions between neighboring molecules when they are in random up-down positions in the N$^*$ phase and when the head-tail orientation equivalence is broken in the N$^*_F$ phase. On approaching the N* phase, the OPR value slightly increases in the N$^*_F$ phase, indicating a small helix elongation before changing its sign in the N* phase.

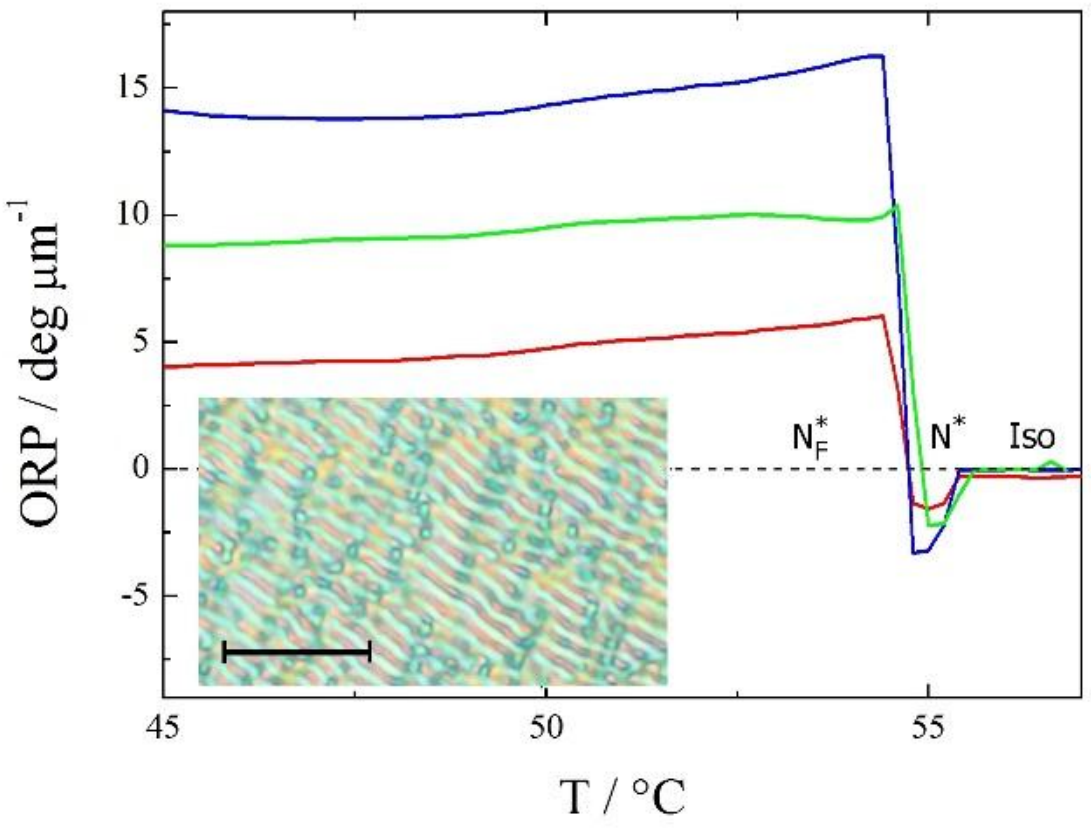

**Figure 2.** Optical rotatory power (ORP) measured in a cell treated for planar anchoring, with blue (488 nm), green (532 nm) and red (690 nm) light for compound **I-4***. The change in the sign of OPR at the N$^*$-N$^*_F$ transition indicates a change of the helix sense. In the inset, the fingerprint texture of the N$^*_F$ phase, the scale bar corresponds to 20 μm.

The spontaneous electric polarization was measured for both, chiral and non-chiral analogues, however due to recrystallization of the sample in the case of the non-chiral material **I-4** it was not possible to record full temperature dependence. Nevertheless, several degrees below the transition to the N$_F$ phase, the values of electric polarization were found comparable for both compounds: ~2.8 mC cm$^{-2}$ and ~2.5 mC cm$^{-2}$ for **I-4** and **I-4*,** respectively. For the chiral material a critical decrease of its $P_s$ value on heating towards the N* phase was observed (Fig. S6).

Dielectric spectroscopy measurements gave information on the polarization vector fluctuations, and were performed in glass cells with transparent ITO or gold electrodes, without polymer aligning layers to avoid the effect of polymer layer capacitance [8] given that the capacitance of the thin polymer layer becomes comparable to the capacitance of the tested compound layer for materials with giant dielectric permittivity, as well as to avoid the formation of twist states induced by interactions with the rubbed surfaces in the N$_F$ phase. In the N phase above the ferroelectric phase, a weak dielectric mode is found, the strength of which increases and frequency decreases on cooling, such a softening of the mode is a characteristic of an increasing correlation length of the polar fluctuations in a non-polar (paraelectric) phase [9].

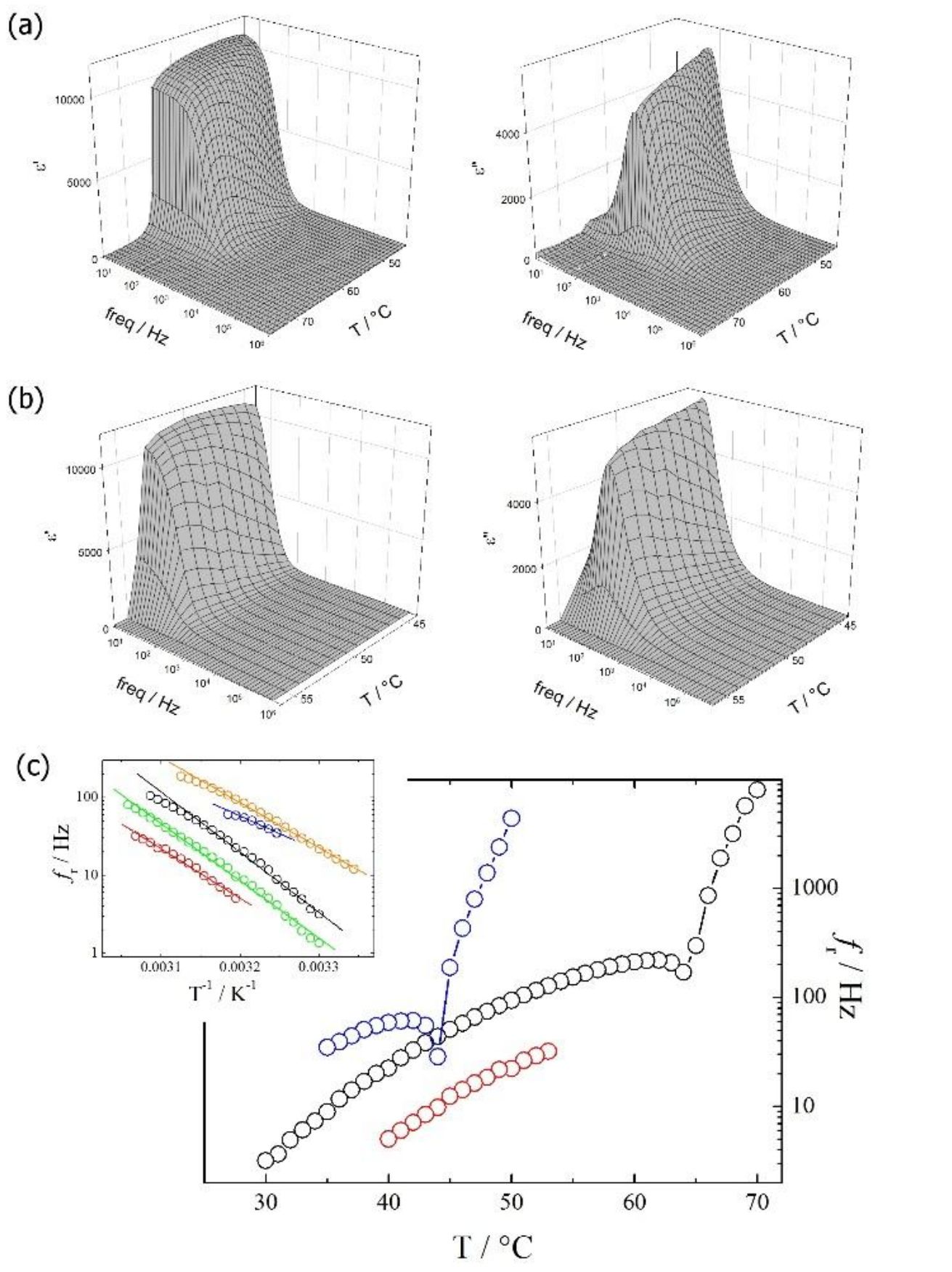

**Figure 3.** Real and imaginary part of dielectric permittivity measured vs. temperature and frequency, in 10-μm-thick cells for (a) **I-4** and (b) **I-4***. (c) Temperature dependence of relaxation frequency for **I-4** (black), **I-4*** (red) and **II-6** (blue). In the inset, 1/T dependence of relaxation frequency in the $N_F$ phase a few degrees below the phase transition from the N phase; comparable slopes of the curves show similar activation energies of fluctuations for the studied materials. It is 148 kJ/mol*K for **I-4** (black), 121 kJ/mol*K for **I-4*** (red), 87 kJ/mol*K for **I-6** (blue), 110 kJ/mol*K for **II-6** (orange), and 141 kJ/mol*K for **II-4*** (green).

For all the compounds in the $N_F$ phase, dielectric spectroscopy shows a single, very strong dielectric relaxation mode (Fig. 3, Figs. S7-S11) which results from a combination of amplitude fluctuation (*i.e.* collective changes of the magnitude of the polarization vector that are pronounced close to the $N_F$-N phase transition) and phase fluctuations (*i.e.*, collective changes of the direction of the polarization vector). The relaxation frequency of this mode depends strongly on the material, namely on the length of the lateral chain: a few K below the N-$N_F$ phase transition, it is ~10 kHz for m=1 [8] and m=2 (**I-1** and **I-2**), ~1kHz for m=3 (**I-3**), ~250 Hz for m=4 (**I-4**), 100 Hz for m=5 (**I-5**), and ~50 Hz for m=6 (**I-6**). As the temperature is lowered, a clear cross over is found from a critical decrease, predominant near the transition to the paraelectric nematic phase, to a decrease driven by non-critical Arrhenius type behavior, $f_r \sim \exp(-E_a/kT)$ deep in the $N_F$ phase [9]. Far from the phase transition temperature the relaxation frequency decrease is mainly due to the increasing rotational viscosity of polarization vector fluctuations. The activation energy for ferroelectric fluctuations in the $N_F$ phase was determined from the relaxation frequency vs. 1/T dependence (inset Fig. 3c), and is comparable (~ 100 kJ/mol K) for all the compounds.

When comparing the dielectric properties of the chiral versus achiral materials, the relaxation frequency in the achiral compound (**I-4**) is slightly larger than in its chiral counterpart (**I-4***), but the temperature dependence of the mode parameters is very similar. In both compounds, the fluctuations could be easily suppressed by a weak bias field, just ~0.2 V/μm is sufficient. For chiral compound **II-4***, in which the additional F atom was introduced and a direct Iso-$N_F$* transition observed, the dielectric mode relaxation frequency is slightly higher than in its chiral non-substituted counterpart **I-4***.

In general, the presence of the chiral center in the lateral substituent caused only a small difference in the dielectric mode frequency in the $N_F$ phase. No pronounced influence of the helical structure on the phason fluctuations were detected for the chiral material. This is in contrast to what is usually reported for chiral smectics, for which the elongation of the helix pitch significantly decreases the relaxation frequency of the phason mode [10].

In addition, second harmonic generation (SHG) measurements were performed to confirm the existence of polar order (Fig. 4).

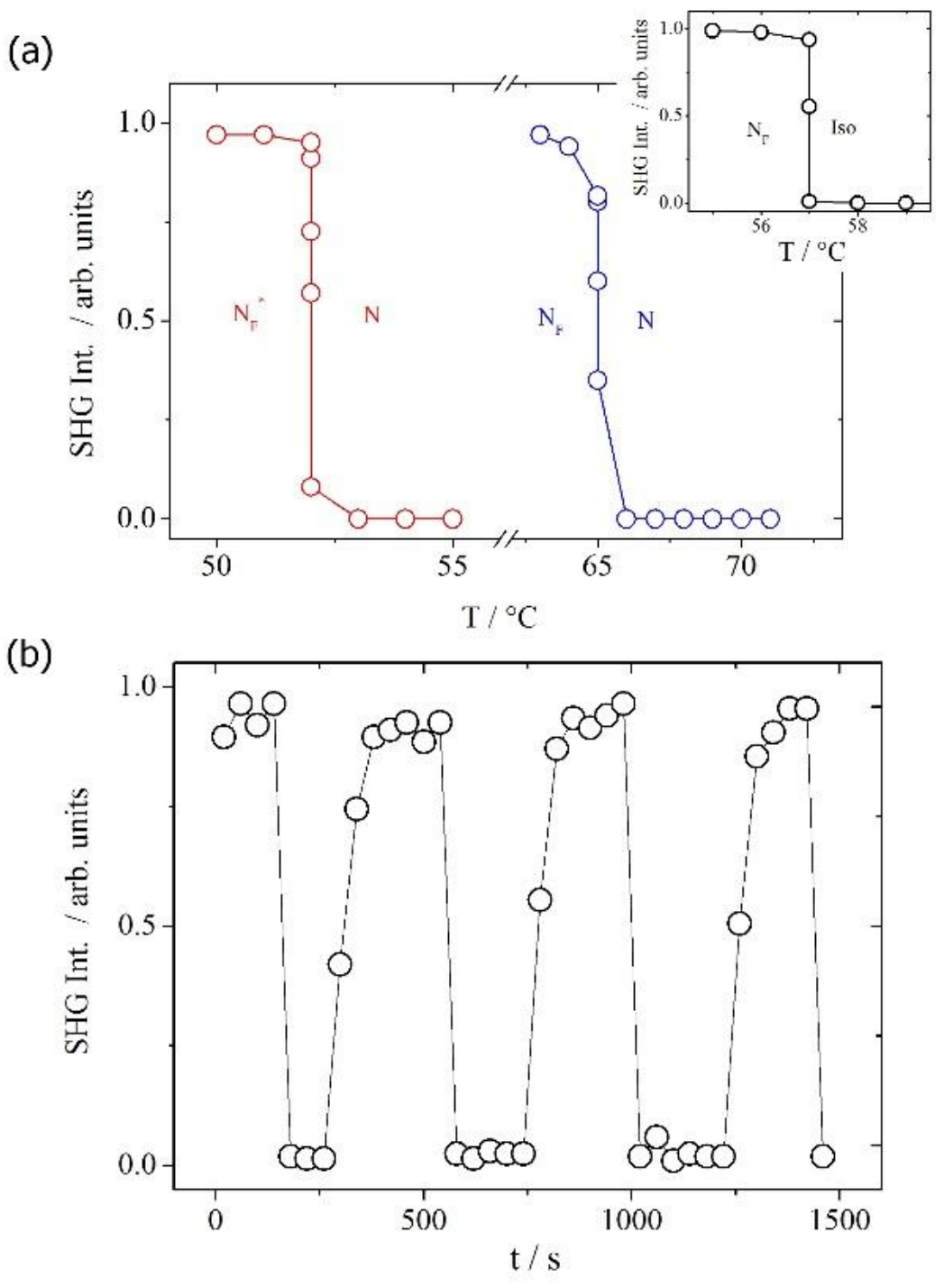

**Figure 4.** Relative intensity of SHG signal versus temperature for (a) **I-4** (blue) and **I-4*** (red) compounds; in the inset, data for compound **II-4*** which shows an Iso-$N^*_F$ transition. (b) The modulation of the SHG signal intensity in a planar 5-μm-thick cell for compound **I-4*** in the $N^*_F$ phase upon application of an electric field; a high intensity SHG signal is observed in the switched-off state and low intensity in the switched on-state, due to the orientation of molecules along the light propagation direction.

The N phase and crystal phase were SHG silent, and no signal was also observed in the N* phase despite it being allowed by the non-centrosymmetric helical structure of the cholesteric phase. In contrast, in the $N_F$ and $N_F$* phases, a strong SHG signal was found (Fig. 4a). It would appear that the detected SHG signal originates from the long-range ordering of the electric dipoles. For materials showing a direct $N_F$-Iso phase transition, the appearance of the signals is abrupt. For both **I-4** and **I-4***, the SHG signal was sufficiently strong to be detected with the naked eye (upon irradiation with λ=1064nm, pulse duration 9 ns and ~2 mJ power in the pulse) in relatively thin, 20 μm cells, and this is not usually the case for liquid crystals. In an external electric field, which causes the reorientation of the electric polarization in the direction of light propagation, the SHG signal disappears, and switching off the field restores the SHG signal, as in the cell with strong planar anchoring, the polarization vector re-adopts the direction perpendicular to the light propagation direction, in $N_F$* phase helix is restored (Fig. 4b).

In summary: intrinsically chiral compounds that form the ferroelectric nematic phase were investigated and their properties compared with those of their achiral counterparts. The chiral center was introduced into a lateral alkyl chain, at the third atom from the mesogenic core. It is known that the further the chiral center is from the mesogenic core, the less the transition temperatures are reduced, however the weaker is the twisting ability of the compound, due to the larger conformational freedom of the molecule. For the compounds studied here, this molecular modification decreased the transition temperatures, resulting in a lower temperature range in which the N and $N_F$ phases are observed than for non-branched compounds. The helical structure in the N* and $N_F$* phases have a few-micron-long pitch, but interestingly with opposite twist sense in these phases. The dielectric relaxation properties were mainly influenced by the rotational viscosity of the material such that lengthening and branching the lateral chain makes the fluctuations of the spontaneous polarization vector slower. The dielectric relaxation mode frequency for the chiral compounds was systematically lower by 50 Hz than for its achiral non-branched counterparts with the same lateral chain length, the activation energy for phason fluctuations was similar for all compounds.

## Acknowledgements

The research was supported by the National Science Centre (Poland) under the grant no. 016/22/A/ST5/00319. C.T.I. and J.M.D.S. acknowledge the financial support of the Engineering and Physical Sciences Research Council [EP/V048775/1].

**Keywords:** nematic • ferroelectric • chirality

Supporting Information

# Intrinsically chiral ferronematic liquid crystals

D. Pociecha,[a] R. Walker,[b] E. Cruickshank,[b] J. Szydlowska,[a] P. Rybak,[a] A. Makal,[a] J. Matraszek,[a] J. M. Wolska,[a] J.M.D. Storey,[b] and C.T. Imrie.[b], E. Gorecka,*[a]

[a] Department of Chemistry, University of Warsaw, Zwirki I Wigury 101, 02-089 Warsaw, Poland
[b] Department of Chemistry, University of Aberdeen, Old Aberdeen AB24 3UE, U.K.

**Table of contents:**

## 1. Materials

Presented reactions were carried out under an argon atmosphere with using a magnetic stirring hotplate. All products were purified by column chromatography with Merck silica gel 60 (230-400 mesh). Analytical thin-layer chromatography (TLC) was performed using Merck Silica Gel 60 Å F254 pre-coated alumina plates (0.25 mm thickness) and visualized using UV lamp (254 nm) and iodine vapor. For column chromatography, the separations were carried out using silica gel grade 60 Å, 40-63 µm particle size and using an appropriate solvent system. During the synthesis following solvents of p.a. quality were used: chloroform, dichloromethane, hexane, toluene, tetrahydrofuran, ethanol and methanol. As a substrates were used Sigma-Aldrich or TCI products without further purification. Presented yields refer to chromatographically and spectroscopically ($^1$H NMR) homogeneous materials.

### Synthesis and characterization of obtained compounds

The materials were synthesised according to previously described procedure [1] presented on Scheme 1.
Into methyl 3-hydroxy-4-methoxybenzoate appropriate alkoxy chains were attached via Williamson reaction. Next the ester group was hydrolysed and transformed into an acid chloride derivative which was substituted by benzyl 4-hydroxybenzoate group via esterification reaction. The protecting group was removed and 4-nitrophenol was attached via the Steglich esterification reaction to yield target compounds. Some compounds were obtained in slightly modified procedure outlined in Scheme 2. The ester bond in intermediate compound 3 was formed by Steglich esterification instead of using acid chloride as in pathway 1.

Following abbreviations are used:
DCC - N,N'-Dicyclohexylcarbodiimide
DMAP - 4-Dimethylaminopyridine

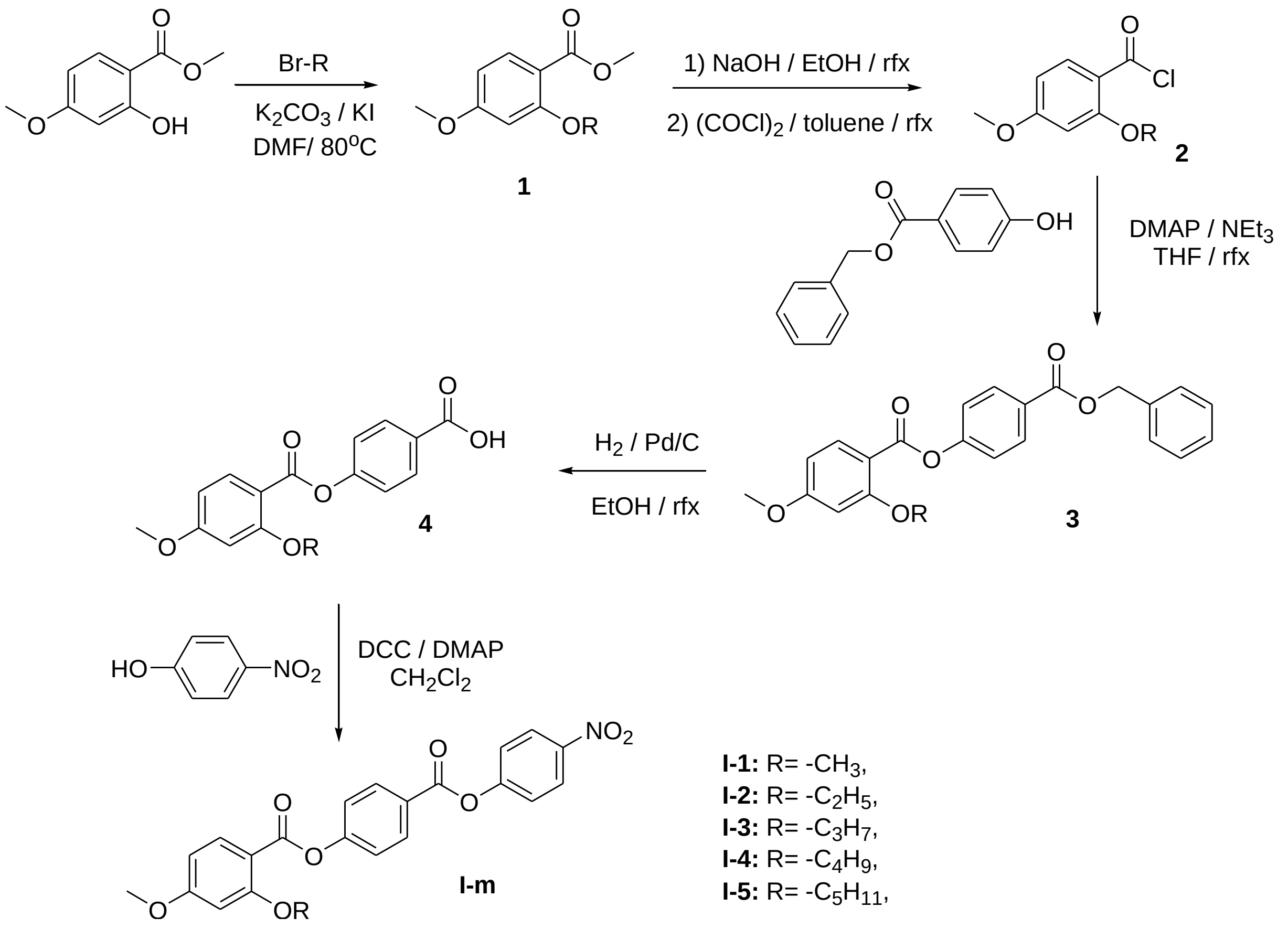

*Scheme 1. Synthesis of the compounds series **I-m***

**Analytical data of obtained materials:**

The intermediate products were characterised by $^{1}$HNMR spectroscopy. For the final compounds $^{1}$HNMR and $^{13}$CNMR spectra were recorded, and were in good agreement with the expected structures for all materials.

**Compounds 1**

*a) Methyl 2,4-dimethoxybenzoate (R=-$CH_3$)*

$^{1}$H NMR: (300MHz, $CDCl_3$) δ: 7.83 (1H; d; J=9.0Hz); 6.49-6.44 (2H; m); 4.00 (3H; s); 3.84 (3H; s); 3.83 (3H; s); Yield: 86%

*b) Methyl 2-ethoxy-4-methoxybenzoate (R=-$C_2H_5$)*

$^{1}$H NMR: (300MHz, $CDCl_3$) δ: 7.83 (1H; d; J=9.3Hz); 6.52-6.44 (2H; m); 4.09 (2H; q; J=6.9Hz); 3.85 (3H; s); 3.83 (3H; s); 1.47 (3H; t; J=6.9Hz); Yield: 82%

*c) Methyl 4-mthoxy-2-propoxybenzoate (R=-$C_3H_7$)*

$^1H$ NMR: (300MHz, $CDCl_3$) δ: 7.84 (1H; d; J=9.0Hz); 6.52-6.43 (2H; m); 3.97 (2H; t; J=6.3Hz); 3.85 (3H; s); 3.83 (3H; s); 1.94-1.78 (2H; m); 1.08 (3H; t; J=7.5Hz); Yield: 87%

*d) Methyl 2-butoxy-4-mthoxy-benzoate (R=-$C_4H_9$)*
$^1H$ NMR: (300MHz, $CDCl_3$) δ: 7.83 (1H; d; J=9.2Hz); 6.55-6.46 (2H; m); 4.01 (2H; t; J=6.2Hz); 3.84 (3H; s); 3.83 (3H; s); 1.87-1.78 (2H; m); 1.60-1.48 (2H; m); 0.98 (3H; t; J=7.3Hz); Yield: 91%

*e) Methyl 4-mthoxy-2-pentoxybenzoate (R=-$C_5H_{11}$)*
$^1H$ NMR: (300MHz, $CDCl_3$) δ: 7.83 (1H; d; J=9.2Hz); 6.50-6.45 (2H; m); 4.00 (2H; t; J=6.5Hz); 3.85 (3H; s); 3.82 (3H; s); 1.91-1.82 (2H; m); 1.52-1.34 (4H; m); 0.95 (3H; t; J=7.0Hz); Yield: 90%

**Compounds 3**

*a) 4-((Benzyloxy)carbonyl)phenyl-2,4-dimethoxybenzoate (R=-$CH_3$)*
$^1H$ NMR: (300MHz, $CDCl_3$) δ: 8.12 (2H; d; J=9.0Hz); 8.07 (1H; d; J=8.5Hz); 7.45-7.34 (5H; m); 7.27 (2H; d; J=9.0Hz); 6.57 (1H; dd; J=8.7Hz; J=2.2Hz); 6.51 (1H; d; J=2.2Hz); 5.37 (2H; s); 3.92 (3H; s); 3.89 (3H; s); Yield: 71%

*b) 4-((Benzyloxy)carbonyl)phenyl-2-ethoxy-4-methoxybenzoate (R=-$C_2H_5$)*
$^1H$ NMR: (300MHz, $CDCl_3$) δ: 8.13 (2H; d; J=9.0Hz); 8.04 (1H; d; J=8.7Hz); 7.48-7.33 (5H; m); 7.28 (2H; d; J=9.0Hz); 6.55 (1H; dd; J=8.7Hz; J=2.4Hz); 6.51 (1H; d; J=2.4Hz); 5.37 (2H; s); 4.12 (2H; q; J=7.2Hz); 3.88 (3H; s); 1.47 (3H; t; J=6.9Hz); Yield: 70%

*c) 4-((Benzyloxy)carbonyl)phenyl-4-methoxy-2-propoxybenzoate (R=-$C_3H_7$)*
$^1H$ NMR: (300MHz, $CDCl_3$) δ: 8.13 (2H; d; J=9.0Hz); 8.03 (1H; d; J=8.7Hz); 7.48-7.32 (5H; m); 7.28 (2H; d; J=9.0Hz); 6.54 (1H; dd; J=8.7Hz; J=2.4Hz); 6.51 (1H; d; J=2.4Hz); 5.37 (2H; s); 4.00 (2H; t; J=6.3Hz); 3.86 (3H; s); 1.93-1.78 (2H; m); 1.05 (3H; t; J=7.5Hz); Yield: 68%

*d) 4-((Benzyloxy)carbonyl)phenyl-2-butoxy-4-methoxybenzoate (R=-$C_4H_9$)*
$^1H$ NMR: (300MHz, $CDCl_3$) δ: 8.13 (2H; d; J=9.2 Hz); 8.03 (1H; d; J=8.8Hz); 7.47-7.34 (5H; m); 7.25 (2H; d; J=9.1Hz); 6.56 (1H; dd; J=8.8Hz; J=2.2Hz); 6.51 (1H; d; J=2.2Hz); 5.37 (2H; s); 4.04 (2H; t; J=7.0Hz); 3.87 (3H; s); 1.87-1.77 (2H; m); 1.58-1.46 (2H; m); 0.97 (3H; t; J=7.5Hz); Yield: 60%

*e) 4-((Benzyloxy)carbonyl)phenyl-4-methoxy-2-pentoxybenzoate (R=-$C_5H_{11}$)*
$^1H$ NMR: (300MHz, $CDCl_3$) δ: 8.14 (2H; d; J=9.0Hz); 8.03 (1H; d; J=8.5Hz); 7.46-7.34 (5H; m); 7.25 (2H; d; J=9.0Hz); 6.55 (1H; dd; J=8.9Hz; J=2.2Hz); 6.51 (1H; d; J=2.2Hz); 5.37 (2H; s); 4.04 (2H; t; J=7.1Hz); 3.88 (3H; s); 1.86-1.79 (2H; m); 1.52-1.23 (4H; m); 0.87 (3H; t; J=7.7Hz); Yield: 67%

**Compounds 4**

*a) 4-(2,4-Dimethoxybenzoyloxy)benzoic acid (R=-$CH_3$)*
$^1H$ NMR: (300MHz, acetone) δ: 8.12 (2H; d; J=8.8Hz); 8.00 (1H; d; J=8.7Hz); 7.34 (2H; d; J=8.8Hz); 6.70 (1H; dd; J=8.7Hz; J=2.2Hz); 6.64 (1H; d; J=2.2Hz); 3.93 (3H; s); 3.92 (3H; s); 1.47 (3H; t; J=6.9Hz); Yield: 61%

*b) 4-(2-Ethoxy-4-methoxybenzoyloxy)benzoic acid (R=-$C_2H_5$)*
$^1H$ NMR: (300MHz, $CDCl_3$) δ: 8.13 (2H; d; J=8.7Hz); 8.04 (1H; d; J=8.7Hz); 7.24 (2H; d; J=8.7Hz); 6.55 (1H; dd; J=8.7Hz; J=2.4Hz); 6.51 (1H; d; J=2.4Hz); 4.12 (2H; q; J=7.8Hz); 3.87 (3H; s); 1.47 (3H; t; J=6.9Hz); Yield: 73%

*c) 4-(4-Methoxy-2-propoxy-benzoyloxy)benzoic acid (R=-$C_3H_7$)*

$^1$H NMR: (300MHz, $CDCl_3$) δ: 8.18 (2H; d; J=9.0Hz); 8.05 (1H; d; J=8.7Hz); 7.32 (2H; d; J=9.0Hz); 6.55 (1H; dd; J=8.7Hz; J=2.4Hz); 6.51 (1H; d; J=2.4Hz); 4.01 (2H; t; J=6.6Hz); 3.87 (3H; s); 1.96-1.79 (2H; m); 1.07 (3H; t; J=7.5Hz); Yield: 68%

*d) 4-(2-Butoxy-4-methoxy-benzoyloxy)benzoic acid (R=-$C_4H_9$)*
$^1$H NMR: (300MHz, $CDCl_3$) δ: 8.17 (2H; d; J=9.1Hz); 8.05 (1H; d; J=8.8Hz); 7.31 (2H; d; J=9.1Hz); 6.55 (1H; dd; J=8.8Hz; J=2.2Hz); 6.50 (1H; d; J=2.2Hz); 4.06 (2H; t; J=6.8Hz); 3.88 (3H; s); 1.86-1.79 (2H; m); 1.57-1.49 (2H; m); 0.95 (3H; t; J=7.5Hz); Yield: 75%

*e) 4-(4-Methoxy-2-pentoxy-benzoyloxy)benzoic acid (R=-$C_5H_{11}$)*
$^1$H NMR: (300MHz, $CDCl_3$) δ: 8.18 (2H; d; J=9.0Hz); 8.05 (1H; d; J=8.7Hz); 7.41 (2H; d; J=9.0Hz); 6.55 (1H; dd; J=8.8Hz; J=2.2Hz); 6.50 (1H; d; J=2.2Hz); 4.05 (2H; t; J=6.7Hz); 3.89 (3H; s); 1.88-1.83 (2H; m); 1.50-1.44 (2H; m); 1.39-1.31 (2H; m); 0.90 (3H; t; J=7.6Hz); Yield: 77%

**Compounds I-m**

*a) 4-((4-Nitrophenoxy)carbonyl)phenyl 2,4-dimethoxybenzoate (R=-$CH_3$);* ***I-1***
$^1$H NMR: (300MHz, $CDCl_3$) δ: 8.34 (2H; d; J=9.2Hz); 8.26 (2H; d; J=8.7Hz); 8.08 (1H; d; J=8.7Hz); 7.43 (2H; d; J=9.1Hz); 7.39 (2H; d; J=8.8Hz); 6.56 (1H; dd; J=8.7Hz; J=2.2Hz); 6.51 (1H; d; J=2.2Hz); 4.07 (2H; t; J=6.9Hz); 3.89 (3H; s);
$^{13}$C NMR (75MHz, $CDCl_3$) δ: 165.54; 163.86; 163.20; 162.35; 156.21; 156.00; 145.57; 134.89; 132.05; 125.72; 125.43; 122.89; 122.63; 110.49; 105.17; 99.89; 64.99; 56.20; 55.73; Yield: 45%

*b) 4-((4-Nitrophenoxy)carbonyl)phenyl 2-ethoxy-4-methoxybenzoate (R=-$C_2H_5$);* ***I-2***
$^1$H NMR: (300MHz, $CDCl_3$) δ: 8.33 (2H; d; J=9.3Hz); 8.26 (2H; d; J=8.7Hz); 8.08 (1H; d; J=8.7Hz); 7.43 (2H; d; J=9.3Hz); 7.39 (2H; d; J=8.7Hz); 6.57 (1H; dd; J=8.7Hz; J=2.4Hz); 6.51 (1H; d; J=2.4Hz); 4.14 (2H; q; J=6.9Hz); 3.89 (3H; s); 1.49 (3H; t; J=6.9Hz);
$^{13}$C NMR (75MHz, $CDCl_3$) δ: 165.58; 163.96; 163.37; 162.19; 156.41; 156.04; 145.77; 134.89; 132.25; 125.86; 125.63; 122.99; 122.83; 111.00; 105.37; 100.23; 64.99; 55.93; 15.00; Yield: 50%

*c) 4-((4-Nitrophenoxy)carbonyl)phenyl 4-methoxy-2-propoxybenzoate (R=-$C_3H_7$);* ***I-3***
$^1$H NMR: (300MHz, $CDCl_3$) δ: 8.33 (2H; d; J=9.3 Hz); 8.26 (2H; d; J=9.0 Hz); 8.06 (1H; d; J=8.7Hz); 7.42 (2H; d; J=9.3Hz); 7.39 (2H; d; J=9.0Hz); 6.56 (1H; dd; J=8.7Hz; J=2.4Hz); 6.51 (1H; d; J=2.4Hz); 4.03 (2H; t; J=6.3Hz); 3.89 (3H; s); 1.96-1.81 (2H; m); 1.07 (3H; t; J=7.5Hz);
$^{13}$C NMR (75MHz, $CDCl_3$) δ: 165.59; 163.97; 163.56; 162.28; 156.45; 156.05; 145.77; 134.98; 132.29; 125.86; 125.64; 123.00; 122.83; 111.99; 105.33; 100.05; 70.77; 55.94; 22.86; 10.96; Yield: 48%

*d) 4-((4-Nitrophenoxy)carbonyl)phenyl 2-butoxy-4-methoxybenzoate (R=-$C_4H_9$);* ***I-4***
$^1$H NMR: (300MHz, $CDCl_3$) δ: 8.33 (2H; d; J=9.1Hz); 8.25 (2H; d; J=9.0 Hz); 8.06 (1H; d; J=8.8Hz); 7.43 (2H; d; J=9.0Hz); 7.37 (2H; d; J=9.0Hz); 6.56 (1H; dd; J=8.8; J=2.2 Hz); 6.51 (1H; d; J=2.2Hz); 4.03 (2H; t; J=6.5Hz); 3.90 (3H; s); 1.94-1.84 (2H; m); 1.60-1.48 (2H; m); 1.00 (3H; t; J=7.6Hz);
$^{13}$C NMR (75MHz, $CDCl_3$) δ: 165.61; 163.91; 163.45; 162.62; 156.19; 156.04; 145.82; 134.79; 131.95; 125.91; 125.63; 122.75; 122.55; 112.02; 105.55; 99.73; 68.57; 56.21; 31.20; 19.31; 14.00; Yield: 43%

*e) 4-((4-Nitrophenoxy)carbonyl)phenyl 4-methoxy-2-pentoxybenzoate (R=-$C_5H_{11}$);* ***I-5***
$^1$H NMR: (300MHz, $CDCl_3$) δ: 8.34 (2H; d; J=9.1Hz); 8.26 (2H; d; J=9.1Hz); 8.06 (1H; d; J=8.7Hz); 7.43 (2H; d; J=9.1Hz); 7.38 (2H; d; J=9.1Hz); 6.56 (1H; dd; J=8.7Hz; J=2.2Hz); 6.52 (1H; d; J=2.2Hz); 4.06 (2H;

t; J=6.7Hz); 3.90 (3H; s); 1.92-1.82 (2H; m); 1.51-1.46 (2H; m); 1.40-1.34 (2H; m); 0.90 (3H; t; J=7.5Hz);
$^{13}C$ NMR (75MHz, $CDCl_3$) δ: 165.60; 163.90; 163.48; 162.22; 156.40; 156.04; 145.78; 134.90; 131.95; 125.86; 125.63; 122.68; 122.50; 111.99; 105.35; 99.73; 68.97; 55.61; 28.81; 28.14; 22.39; 14.00;
Yield: 47%

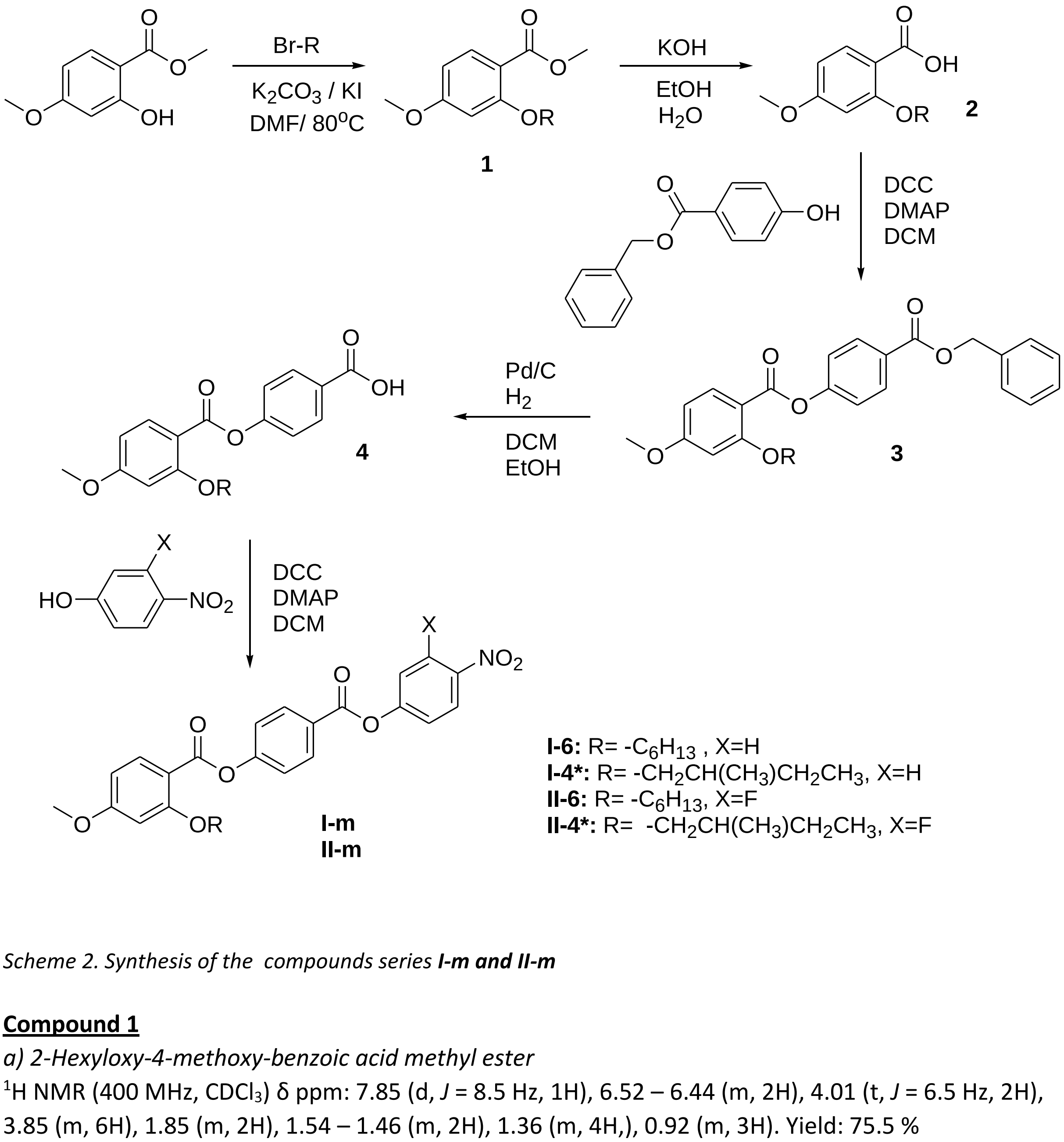

*Scheme 2. Synthesis of the compounds series **I-m and II-m***

**Compound 1**

*a) 2-Hexyloxy-4-methoxy-benzoic acid methyl ester*
$^1H$ NMR (400 MHz, $CDCl_3$) δ ppm: 7.85 (d, *J* = 8.5 Hz, 1H), 6.52 – 6.44 (m, 2H), 4.01 (t, *J* = 6.5 Hz, 2H), 3.85 (m, 6H), 1.85 (m, 2H), 1.54 – 1.46 (m, 2H), 1.36 (m, 4H,), 0.92 (m, 3H). Yield: 75.5 %

*b) (S)-4-Methoxy-2-(2-methyl-butoxy)-benzoic acid methyl ester*
$^1H$ NMR (400 MHz, $CDCl_3$) δ ppm: 7.86 (d, *J* = 8.6 Hz, 1H), 6.52 – 6.42 (m, 2H), 3.94 – 3.81 (m, 8H), 1.92 (m, 1H), 1.70 – 1.55 (m, 1H), 1.41 – 1.20 (m, 1H), 1.08 (d, *J* = 6.7 Hz, 3H), 0.97 (t, *J* = 7.4 Hz, 3H). Yield: 30.1 %

**Compound 2**

*a) 2-Hexyloxy-4-methoxy-benzoic acid*

$^{1}$H NMR (400 MHz, DMSO) δ ppm: 7.46 (d, *J* = 8.6 Hz, 1H), 6.41 – 6.29 (m, 2H), 3.79 (t, *J* = 6.4 Hz), 3.58 (s, 3H), 1.54 – 1.42 (m, 2H), 1.22 (p, *J* = 7.2 Hz, 2H), 1.08 (m, 4H), 0.65 (m, 3H). Yield: 90.1 %

*b) (S)-4-Methoxy-2-(2-methyl-butoxy)-benzoic acid*

$^{1}$H NMR (400 MHz, DMSO) δ ppm: 7.68 (d, *J* = 8.6 Hz, 1H), 6.70 – 6.43 (m, 2H), 3.94 – 3.77 (m, 5H), 1.87 – 1.75 (m, *J* = 6.5 Hz, 1H), 1.55 (m, 1H), 1.25 (m, 1H), 0.99 (d, *J* = 6.7 Hz, 3H), 0.90 (t, *J* = 7.4 Hz, 3H). Yield: 98.1 %

Compound 3

*a) 2-Hexyloxy-4-methoxy-benzoic acid 4-benzyloxycarbonyl-phenyl ester*

$^{1}$H NMR (400 MHz, $CDCl_3$) δ ppm: 8.24 – 8.11 (m, 2H), 8.05 (d, *J* = 8.8 Hz, 1H), 7.55 – 7.21 (m, 7H), 6.62 – 6.47 (m, 2H), 5.40 (s, 2H), 4.06 (t, *J* = 6.5 Hz, 2H), 3.90 (s, 3H), 1.91 – 1.80 (m, 2H), 1.55 – 1.44 (m, 2H), 1.40 – 1.18 (m, 4H), 0.93 – 0.84 (m, 3H). Yield: 73.8 %

*b) (S)-4-Methoxy-2-(2-methyl-butoxy)-benzoic acid 4-benzyloxycarbonyl-phenyl ester*

$^{1}$H NMR (400 MHz, DMSO) δ ppm: 8.08 (d, *J* = 8.6 Hz, 2H), 7.94 (d, *J* = 8.7 Hz, 1H), 7.54 – 7.28 (m, 7H), 6.74 – 6.58 (m, 2H), 5.37 (s, 2H), 3.98 – 3.88 (m, 2H), 3.87 (s, 3H), 1.80 (m, 1H), 1.61 – 1.46 (m, 1H), 1.32 – 1.19 (m, 1H), 0.97 (d, *J* = 6.7 Hz, 3H), 0.86 (t, *J* = 7.5 Hz, 3H). Yield: 67.7 %

**Compound 4**

*a) 2-Hexyloxy-4-methoxy-benzoic acid 4-carboxy-phenyl ester*

$^{1}$H NMR (400 MHz, DMSO) δ ppm: 13.01 (s, 1H), 8.06 – 7.98 (m, 2H), 7.92 (d, *J* = 8.7 Hz, 1H), 7.36 – 7.29 (m, 2H), 6.72 – 6.63 (m, 2H), 4.08 (t, *J* = 6.2 Hz, 2H), 3.87 (s, 3H), 1.72 (p, *J* = 6.6 Hz, 2H), 1.43 (q, *J* = 7.5 Hz, 2H), 1.24 (m, 4H), 0.80 (t, *J* = 6.9 Hz, 3H). Yield: 92.7 %

*b) (S)-4-Methoxy-2-(2-methyl-butoxy)-benzoic acid 4-carboxy-phenyl ester*

$^{1}$H NMR (400 MHz, DMSO) δ ppm: 13.01 (s, 1H), 8.06 – 7.98 (m, 2H), 7.93 (d, *J* = 8.7 Hz, 1H), 7.36 – 7.29 (m, 2H), 6.72 – 6.63 (m, 2H), 3.96 (m, 2H), 3.87 (s, 3H), 1.81 (h, *J* = 6.4 Hz, 1H,), 1.61 – 1.49 (m, 1H), 1.24 (m, 1H), 0.98 (d, *J* = 6.7 Hz, 3H), 0.87 (t, *J* = 7.4 Hz, 3H). Yield: 56.3 %

**Compound I-m**

*a) 2-Hexyloxy-4-methoxy-benzoic acid 4-(4-nitro-phenoxycarbonyl)-phenyl ester, **I-6***

$^{1}$H NMR (400 MHz, DMSO) δ ppm: 8.38 (d, *J* = 8.9 Hz, 2H), 8.25 (d, *J* = 8.3 Hz, 2H), 7.95 (d, *J* = 8.7 Hz, 1H), 7.66 (d, *J* = 8.7 Hz, 2H), 7.47 (d, *J* = 8.3 Hz, 2H), 6.71 – 6.64 (m, 2H), 4.10 (t, *J* = 6.2 Hz, 2H), 3.88 (s, 3H), 1.73 (p, *J* = 6.8 Hz, 2H), 1.45 (p, *J* = 7.3 Hz, 2H), 1.25 (m, 4H), 0.82 (t, *J* = 6.8 Hz, 3H). $^{13}$C NMR (101 MHz, DMSO) δ ppm: 165.42, 163.79, 163.29, 161.68, 156.02, 155.97, 145.69, 134.52, 132.22, 126.03, 125.81, 123.86, 123.16, 110.40, 106.34, 100.08, 68.85, 56.20, 31.35, 29.00, 25.59, 22.51, 14.31. Yield: 14.4 %

*b) (S)-4-Methoxy-2-(2-methyl-butoxy)-benzoic acid 4-(4-nitro-phenoxycarbonyl)-phenyl ester, **I-4****

$^{1}$H NMR (400 MHz, DMSO) δ ppm: 8.43 – 8.32 (m, 2H), 8.29 – 8.22 (m, 2H), 7.96 (d, *J* = 8.7 Hz, 1H), 7.69 – 7.62 (m, 2H), 7.47 (d, *J* = 8.6 Hz, 2H), 6.74 – 6.64 (m, 2H), 3.98 (m, 2H), 3.88 (s, 3H), 1.83 (h, *J* = 6.4 Hz, 1H), 1.63 – 1.48 (m, 1H), 1.26 (m, 1H), 0.99 (d, *J* = 6.7 Hz, 3H), 0.88 (t, *J* = 7.4 Hz, 3H). $^{13}$C NMR (100 MHz, DMSO) δ ppm: 165.47, 163.75, 163.31, 161.81, 156.01, 155.97, 145.65, 134.58, 132.24, 126.03, 125.78, 123.84, 123.15, 110.19, 106.34, 99.86, 73.29, 56.20, 34.70, 25.86, 16.79, 11.65. m/z: $[M+H]^+$ Calculated mass for $C_{26}H_{26}NO_8$: 480.1658, Found: 480.1653. Difference: 1.0 ppm. Yield: 39.8 %

**Compound II-m**

a) 2-Hexyloxy-4-methoxy-benzoic acid 4-(3-fluoro-4-nitro-phenoxycarbonyl)-phenyl ester, ***II-6***

$^{1}H$ NMR (400 MHz, DMSO) δ ppm: 8.33 (t, *J* = 8.9 Hz, 1H), 8.24 (d, *J* = 8.6 Hz, 2H,), 7.96 (d, *J* = 8.7 Hz, 1H), 7.80 (m, 1H), 7.49 (m, 3H), 6.86 – 6.35 (m, 2H), 4.10 (t, *J* = 6.2 Hz, 2H), 3.88 (s, 3H), 1.73 (p, *J* = 6.6 Hz, 2H), 1.44 (m, 2H), 1.29 – 1.19 (m, 4H,), 0.82 (t, *J* = 6.8 Hz, 3H,);
$^{19}F$ NMR (376 MHz, DMSO) δ ppm: -115.37;
$^{13}C$ NMR (100 MHz, DMSO) δ ppm: 165.41, 163.44, 163.24, 161.67, 156.99, 156.10, 155.98, 154.38, 135.24, 135.17, 134.51, 132.28, 128.00, 125.76, 123.18, 119.72, 119.68, 113.43, 113.19, 110.35, 106.33, 100.07, 68.84, 56.20, 31.35, 29.00, 25.59, 22.51, 14.31. Yield: 11.5 %

*b) (S)-4-Methoxy-2-(2-methyl-butoxy)-benzoic acid 4-(3-fluoro-4-nitro-phenoxycarbonyl)-phenyl ester,* ***II-4****
$^{1}H$ NMR (400 MHz, DMSO) δ ppm: 8.33 (t, *J* = 8.8 Hz, 1H), 8.28 – 8.21 (m, 2H), 7.96 (d, *J* = 8.7 Hz, 1H), 7.79 (m, 1H), 7.49 (t, *J* = 9.3 Hz, 3H), 6.76 – 6.59 (m, 2H), 4.04 – 3.91 (m, 2H), 3.88 (s, 3H), 1.83 (q, *J* = 6.5 Hz, 1H), 1.56 (m, 1H), 1.26 (m, 1H,), 0.99 (d, *J* = 6.7 Hz, 3H), 0.88 (t, *J* = 7.4 Hz, 3H);
$^{19}F$ NMR (376 MHz, DMSO) δ ppm: -115.38.
$^{13}C$ NMR (100 MHz, DMSO) δ ppm: δ C (101 MHz, DMSO-d6) 165.49, 163.44, 163.31, 161.81, 156.99, 156.10, 156.00, 154.38, 134.59, 135.25, 135.18, 132.32, 128.01, 125.78, 123.19, 119.73, 119.69, 113.44, 113.20, 110.17, 106.37, 99.87, 73.29, 56.22, 34.69, 25.86, 16.80, 11.66. Yield: 8.7 %

## 3. Experimental methods

***NMR spectra:*** NMR spectra were recorded on a NMR Bruker AVANCE 300 MHz spectrometer (for compounds showed in the Scheme1) or a NMR Bruker Avance III HD 400 MHz spectrometer (for compounds showed in Scheme2) using tetramethylsilane as an internal standard. Chemical shifts are reported in ppm.

***Calorimetric Measurements:*** Calorimetric studies were performed with a TA DSC Q200 calorimeter, 1 - 3 mg samples were sealed in aluminum pans and kept in nitrogen atmosphere during measurement, and both heating and cooling scans were performed with a rate of 5–10 K/min.

***Microscopic Studies*:** Optical studies were performed by using the Zeiss Axio Imager A2m polarized light microscope, equipped with Linkam heating stage. Samples were prepared in commercial cells (AWAT) of various thickness (1.5 – 20 μm) having ITO electrodes and surfactant layer for either planar or homeotropic alignment, HG and HT cells, respectively.

***X-Ray Diffraction:*** The wide angle X-ray diffraction patterns were obtained with the Bruker D8 GADDS system (CuKα radiation, Goebel mirror monochromator, 0.5 mm point beam collimator, Vantec2000 area detector), equipped with modified Linkam heating stage. Samples were prepared as droplets on a heated surface. For single crystal X-ray diffraction measurements a sample was mounted on a kapton loop with a drop of ParatoneN oil. Intensity data were measured on Rigaku Oxford Diffraction Supernova 4 circle diffractometer equipped with copper (CuKα) microsource and Atlas CCD detector at 120K. The temperature of the sample was controlled with a precision of ± 0.1 K using Oxford Cryosystems cooling device. The data were collected, integrated and scaled with CrysAlis1711 software [2]. The structures were solved by direct methods using SXELXT [3] and refined by full-matrix least squares procedure with SHELXL [4] within OLEX2 [5] graphical interface. Structure were deposited with CCDC and can be retrieved upon request.

***Birefringence Measurements:*** Birefringence was measured with a setup based on a photoelastic modulator (PEM-90, Hinds) working at a modulation frequency f =50 kHz; as a light source, a halogen lamp (Hamamatsu LC8) was used, equipped with a narrow bandpass filter (532 nm). The signal from a photodiode (FLC Electronics PIN-20) was de-convoluted with a lock-in amplifier (EG&G 7265) into 1f and 2f components to yield a retardation induced by the sample. Knowing the sample thickness, the retardation was recalculated into optical birefringence. Uniformly aligned samples in HG cells, 1.5- and 3-μm-thick, were measured.

***Optical Rotatory Power measurements*:** ORP was measured using Nikon microscope equipped with rotating polarizer mounted below the sample and polarizer at fixed position above the sample. The intensity of light transmission, passing through the optical setup was measured by photomultiplier mounted on the microscope. The intensity of light was measured as the function of lower polarizer position in the range 0-180 deg, and fitted to the sinusoidal function to find minimum intensity polarizer position.

***Spontaneous Electric Polarization Measurements*:** Values of the spontaneous electric polarization were obtained from the current peaks recorded during Ps switching upon applying triangular voltage at a frequency of 1-10 Hz. The 3 to 20-μm-thick cells with ITO electrodes were used, switching current was determined by recording the voltage drop at the resistivity of 5kΩ in serial connection with the sample. The current peak was integrated over time to calculate the surface electric charge and evaluate polarization value.

***Dielectric spectroscopy:*** The complex dielectric permittivity, $\varepsilon^*$, was measured in 1 Hz – 10 MHz frequency ($f$) range using Solatron 1260 impedance analyzer. Material was placed in glass cells with ITO or Au electrodes (and no polymer alignment layer) and thickness ranging from 1.5 to 20 microns. The relaxation frequency, $f_r$, and dielectric strength of the mode, $\Delta\varepsilon$, were evaluated by fitting complex dielectric permittivity to the Cole-Cole formula: $\varepsilon - \varepsilon_\infty = \sum \frac{\Delta\varepsilon}{(1+\frac{if}{fr})^{1-\alpha}} + i\frac{\delta}{2\pi\varepsilon_o f}$ , where $\varepsilon_\infty$ is high frequency dielectric constant, $\alpha$ is distribution parameter of the mode and $\sigma$ is low frequency conductivity, respectively.

***SHG measurements:*** The SHG response was investigated using solid-state laser EKSPLA NL202. The 9 ns laser pulses at 10 Hz repetition rate and maximal ~2 mJ power in the pulse at λ=1064 nm were applied. Pulse energy was adjusted to each sample to avoid its decomposition. The infra-red beam was incident onto the LC homogenous cell of thickness 1.7-20 µm. IR pass filter was placed at the entrance to the sample and green pass filter at exit of the sample, the emitted SHG radiation was detected using photon counting head (Hamamatsu H7421) with power supply unit (C8137). The signal intensity was estimated by photon counting software connected to the oscilloscope (Agilent Technologies DSO6034A).

## 3. Crystallographic data

### 3.1 Compound I-4

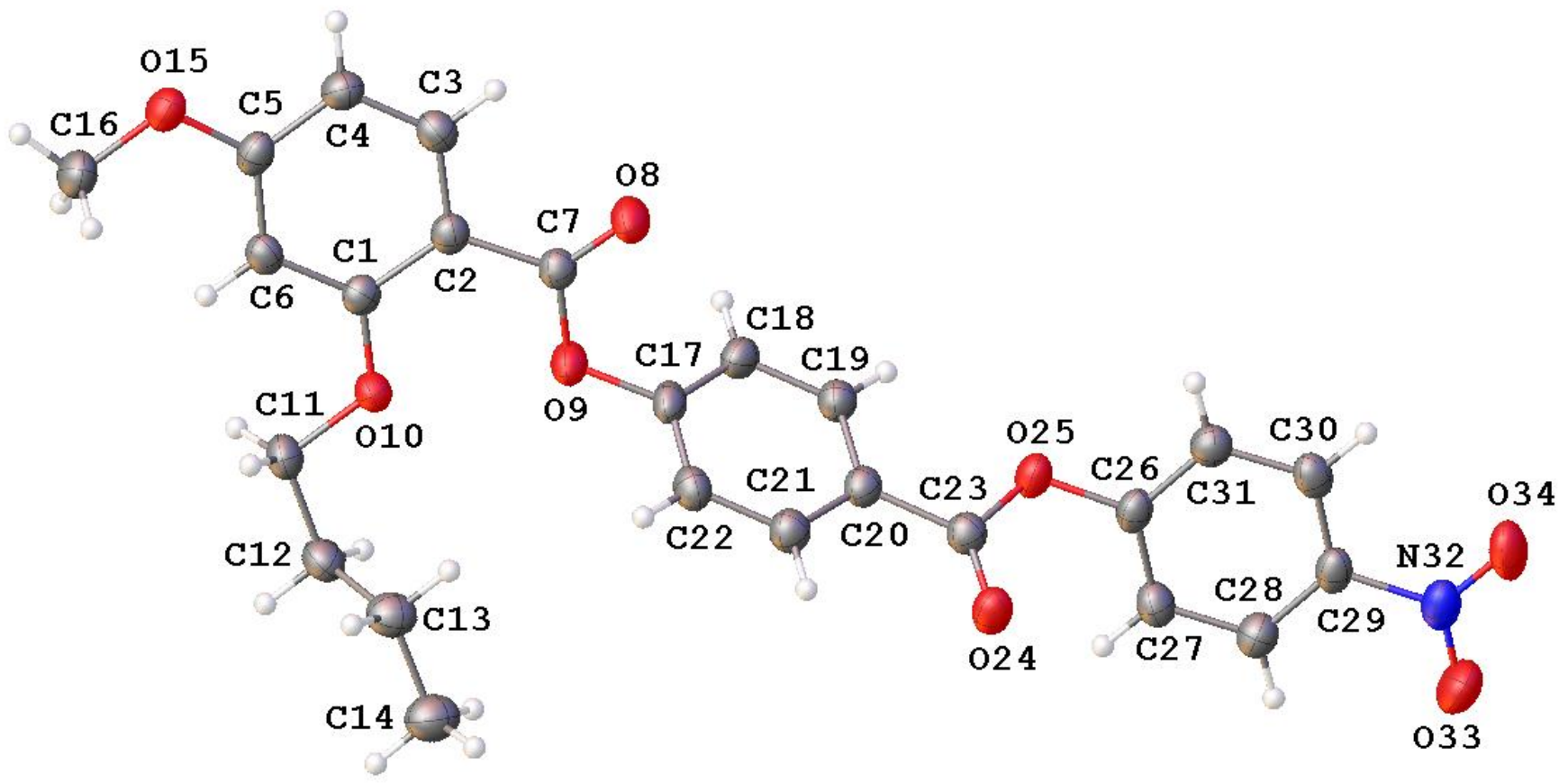

**Figure S1.** Molecular structure of compound **I-4** determined from crystal data and structure refinement, representation with atomic displacement parameters at 50% probability level.

Crystal data and structure refinement for compound **I-4**

Crystal system: monoclinic
Space group: $P2_1/c$
Formula: C25 H23 N O8

**a** = 16.6955(16) Å
**b** = 9.3735(7) Å
**c** = 15.0994(12) Å
**α** = 90°
**β** = 111.798(9)°
**γ** = 90°
**V** = 2194.0(3) Å$^3$
**Z** = 4

There is 1 molecule in the asymmetric unit and 4 molecules in the unit cell, and no traces of solvent in the structure. The molecule displays a single conformation with the lateral aliphatic chain in fully stretched conformation, there is no traces of disorder. The angles between the planes of the consecutive phenyl rings (ring I: C1 – C6, ring II: C18 – C22, ring III: C26 - C30), are about 60 degrees. The molecules are aligned along [2,0,1] crystallographic direction. The molecules related by $\mathbf{c}_{[010]}$ glide plane display π … π stacking of the phenyl rings I with II and II with III. Other intermolecular interactions involve: C – H … O interactions from methoxy group and the aliphatic linker $CH_2$ towards the carbonyl oxygens O8 and O24 C – H … O interactions from phenyl group III towards the oxygens of the nitro group.

The molecules of **I-4** do not form clearly separated layers in the crystal structure.

[100]

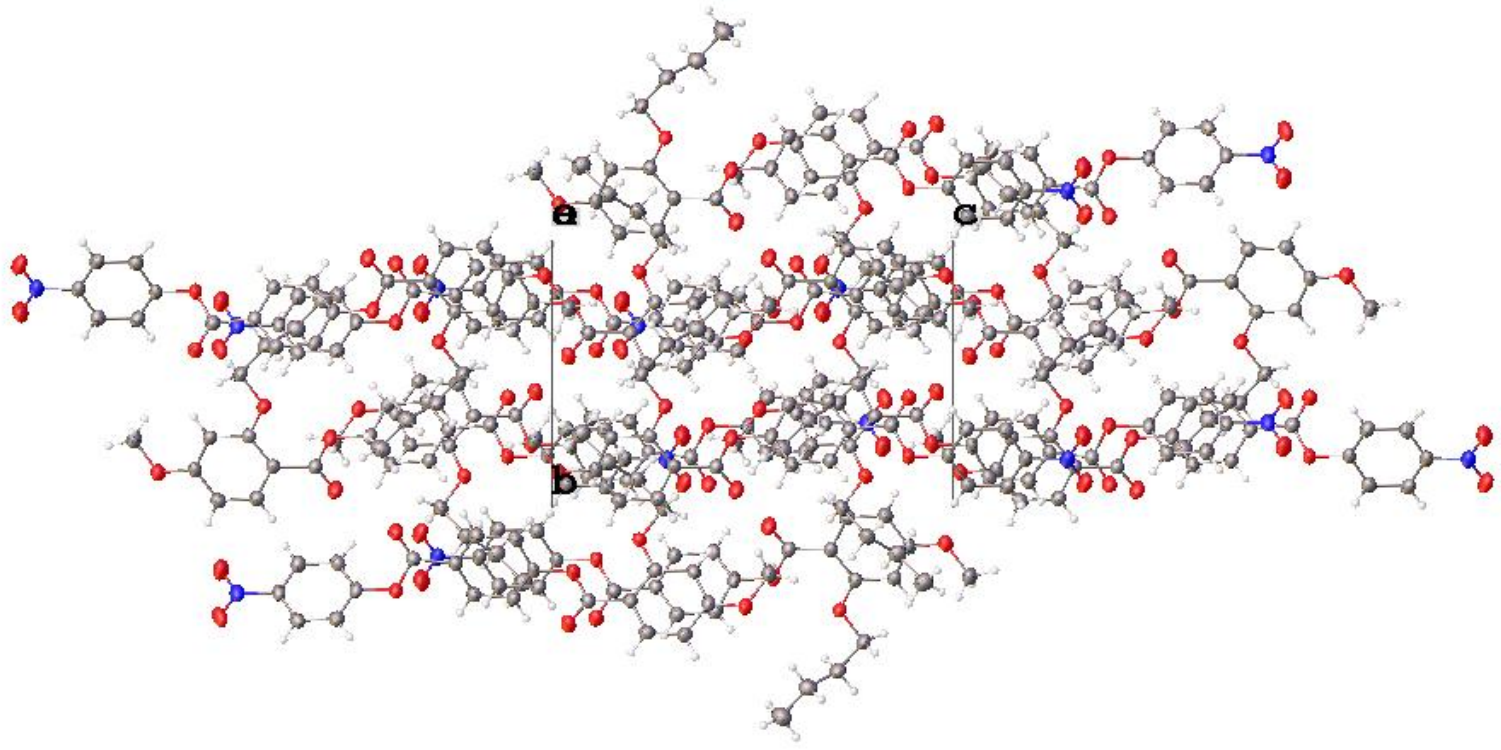

[010]

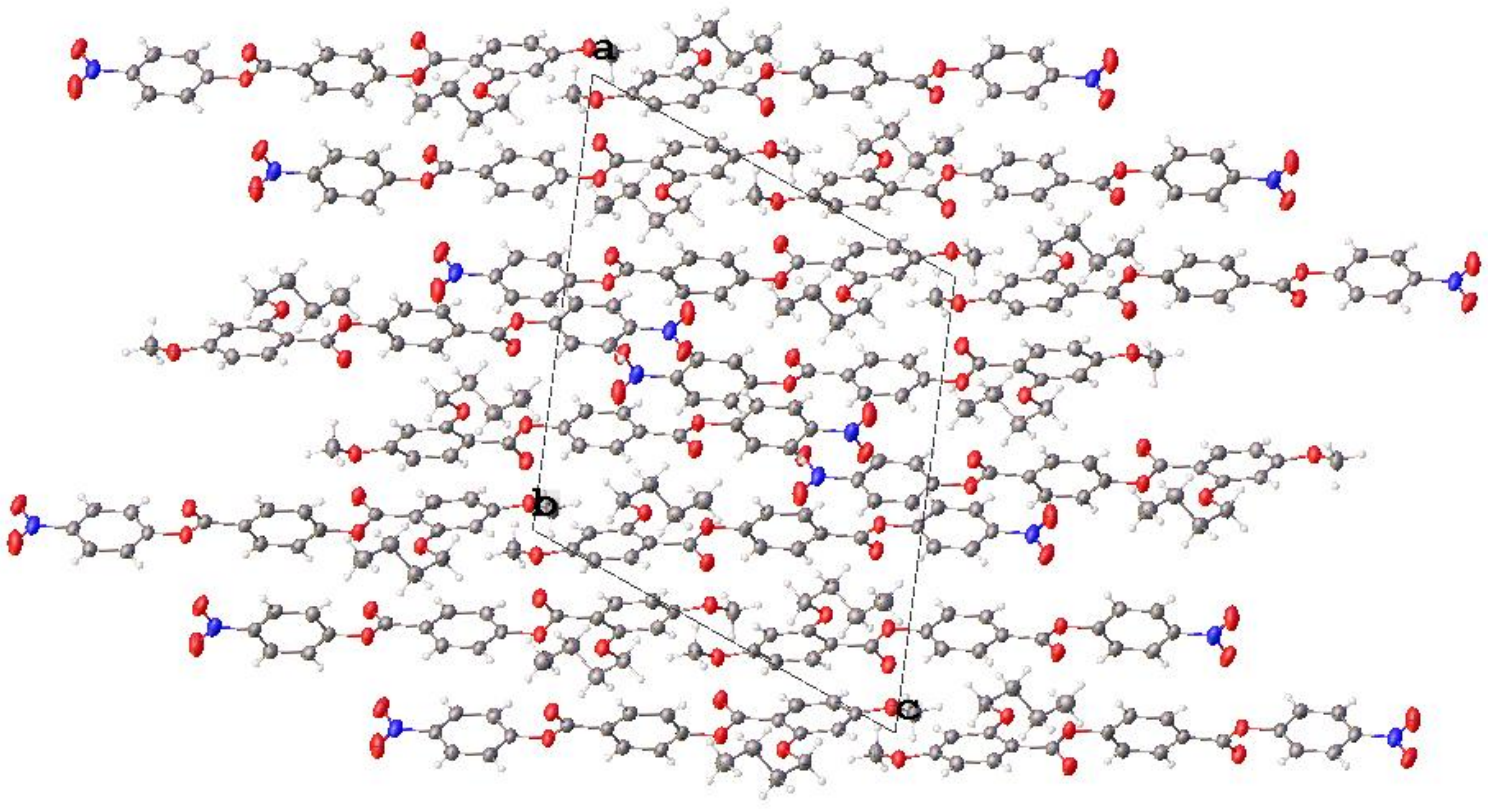

[001]

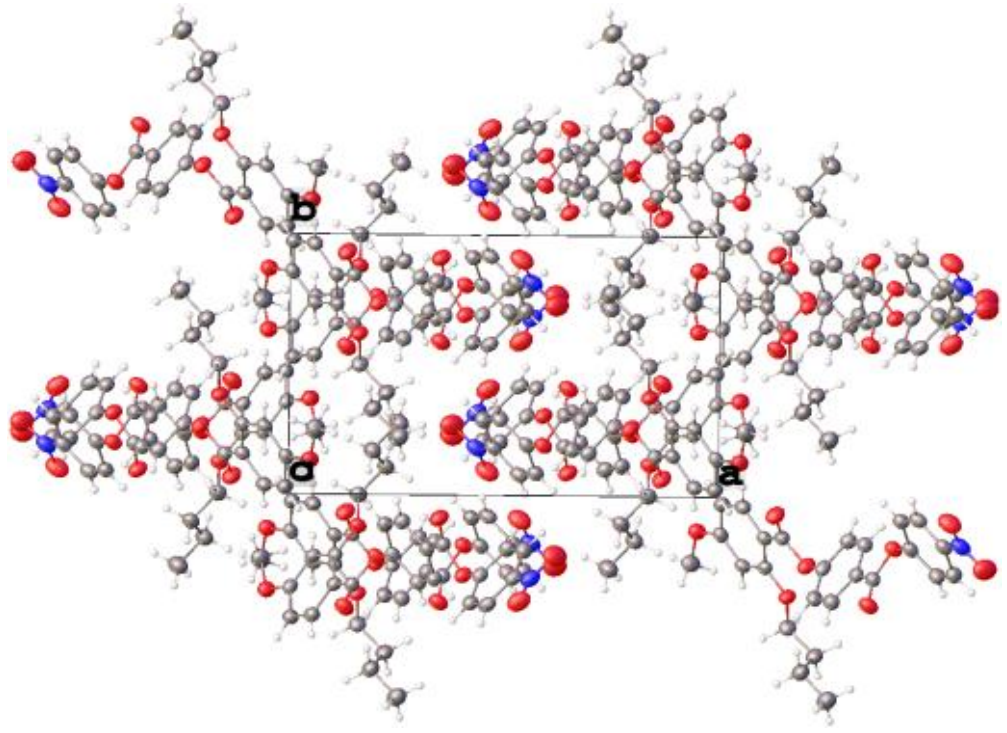

**Figure S2.** Orientation of molecules **I-4** in crystal lattice viewed along [100], [010] and [001] crystallographic directions.

### 3.2 Compound I-3

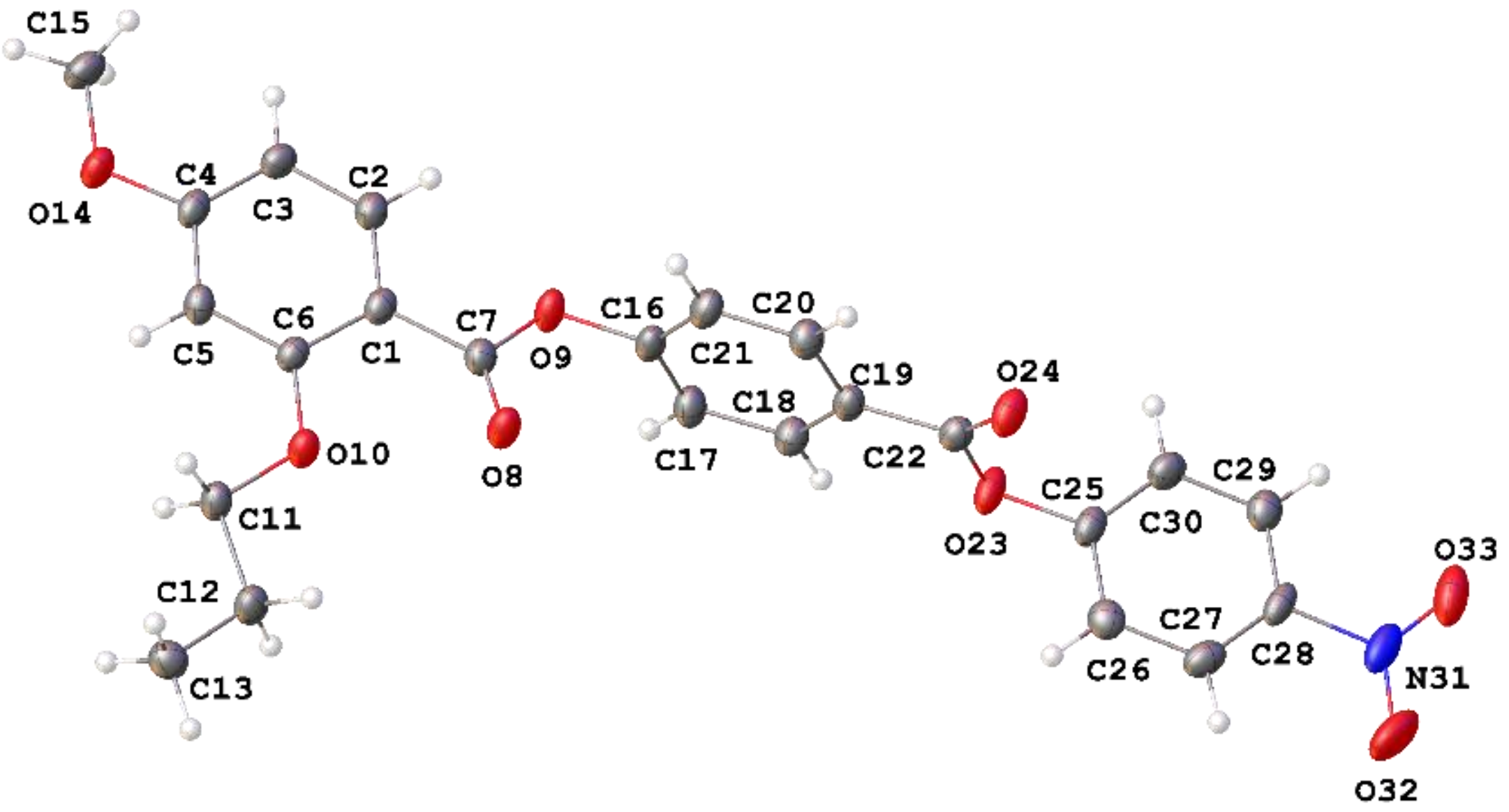

**Figure S3.** Molecular structure of compound **I-3** determined from crystal data and structure refinement, representation with atomic displacement parameters at 50% probability level.

<u>Crystal data and structure refinement for compound **I-3**</u>

| | |
|---|---|
| Crystal system: monoclinic | **a** = 11.0413(7) Å |
| Space group: $P2_1/c$ | **b** = 25.266(2) Å |
| Formula: C24 H21 N O8 | **c** = 7.7947(7) Å |
| | **α** = 90° |
| | **β** = 103.501(7)° |
| | **γ** = 90° |
| | **V** = 2144.4(3) $Å^3$ |
| | **Z** = 4 |

Compound crystallizes in monoclinic system in centrosymmetric space group, with one molecule constituting the asymmetric unit and no solvent. Three aromatic rings are not exactly co-planar, but adopt similar tilt. Major intermolecular interaction is π-stacking of the flat nitrophenyl rings of two antiparallel molecules, as illustrated in Figure S4.

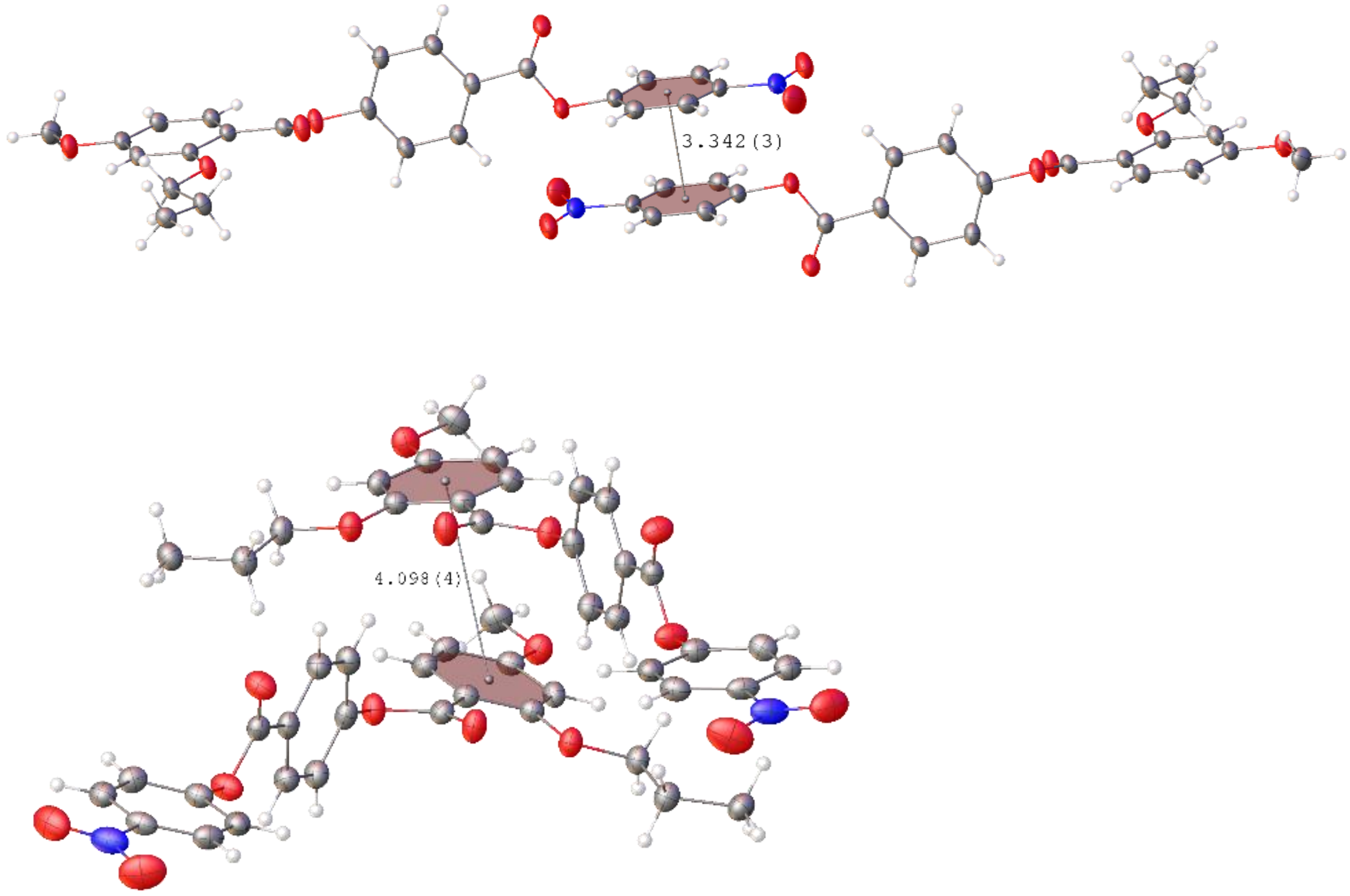

**Figure S4.** The most important intermolecular interactions in crystal structure of compound **I-3.**

## 4. Additional experimental results

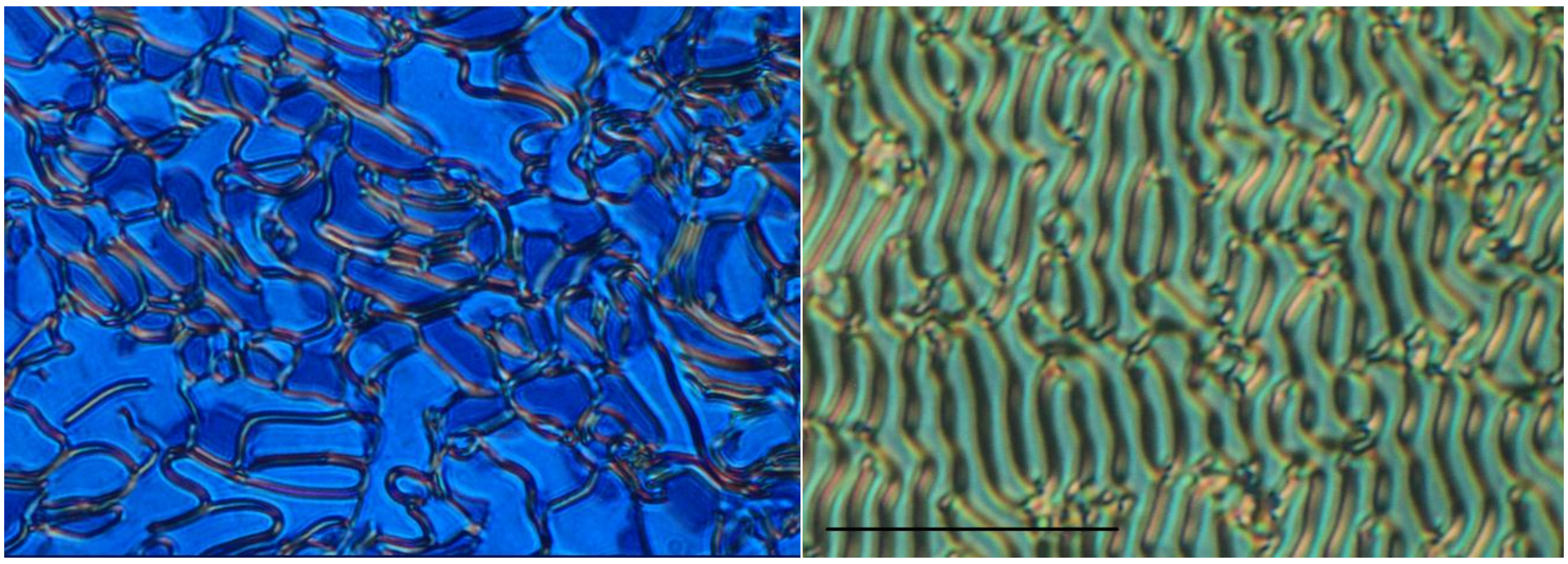

**Figure S5**. Optical textures (crossed polarizers) of N*$_F$ phase for compounds: **I-4*** (left, fast cooled sample) and **II-4*** (right, slowly cooled sample) at room temperature in a 5-μm-thick cell with planar anchoring condition. Scale bar is 40 microns.

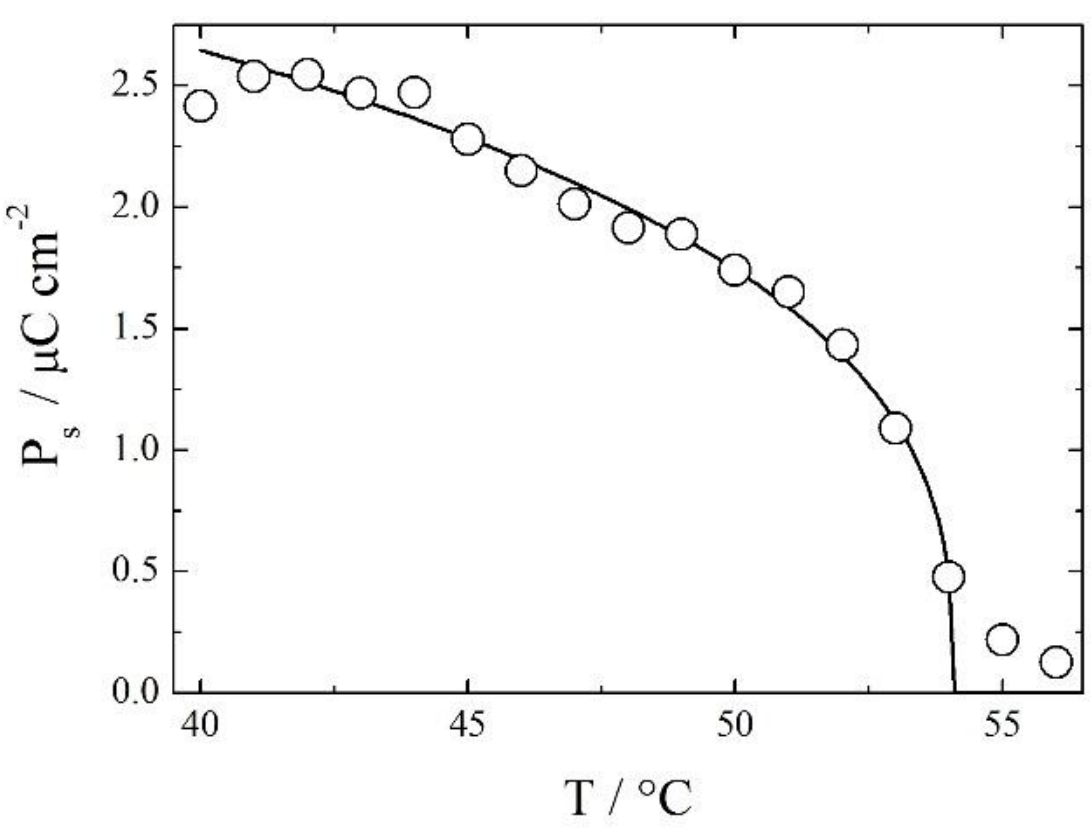

**Figure S6**. Spontaneous electric polarization of N*$_F$ phase vs. temperature for compound: **I-4*.**

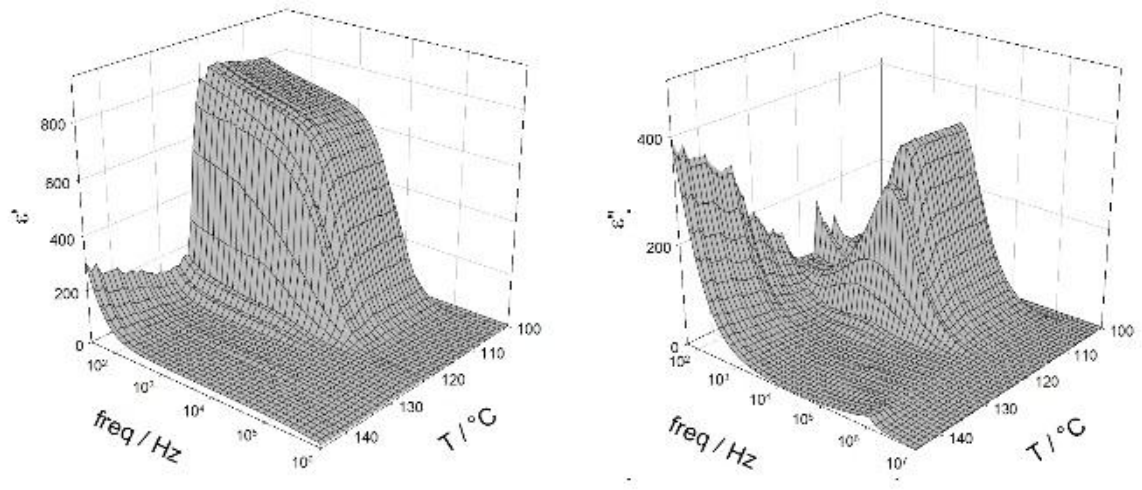

**Figure S7.** Real and imaginary part of dielectric permittivity measured vs. temperature and frequency, in 10-μm-thick cells for **I-2**

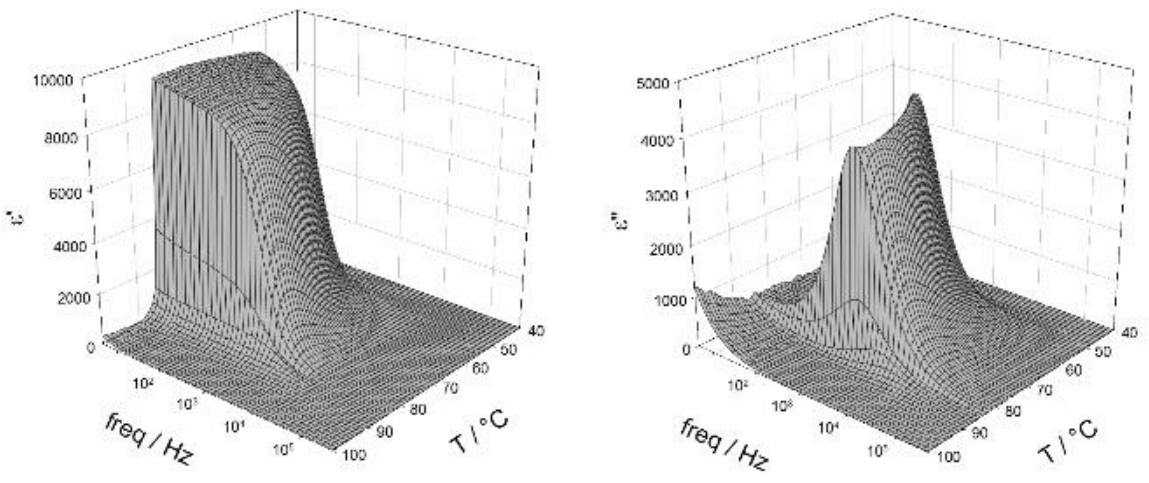

**Figure S8.** Real and imaginary part of dielectric permittivity measured vs. temperature and frequency, in 10-μm-thick cells for **I-3**

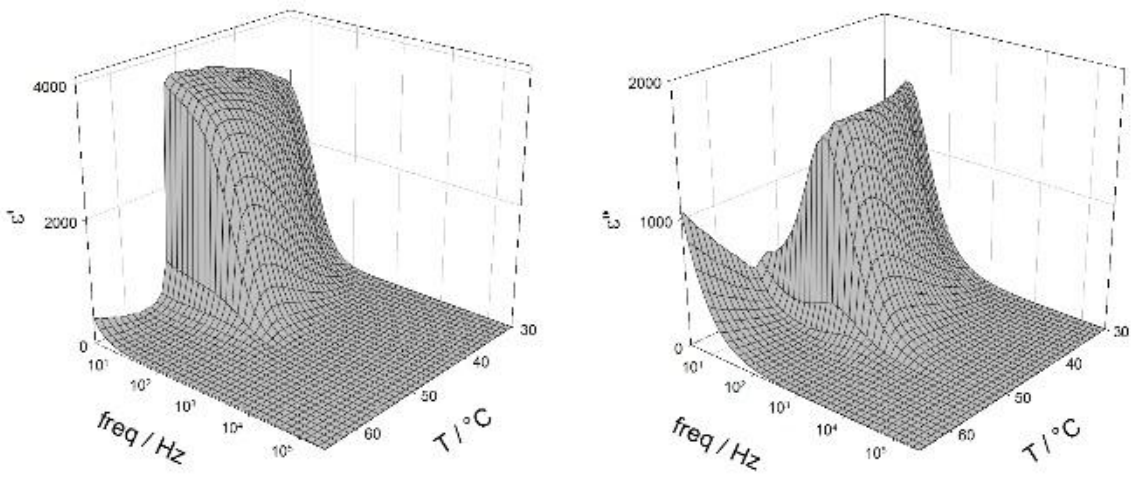

**Figure S9.** Real and imaginary part of dielectric permittivity measured vs. temperature and frequency, in 10-μm-thick cells for **I-5**

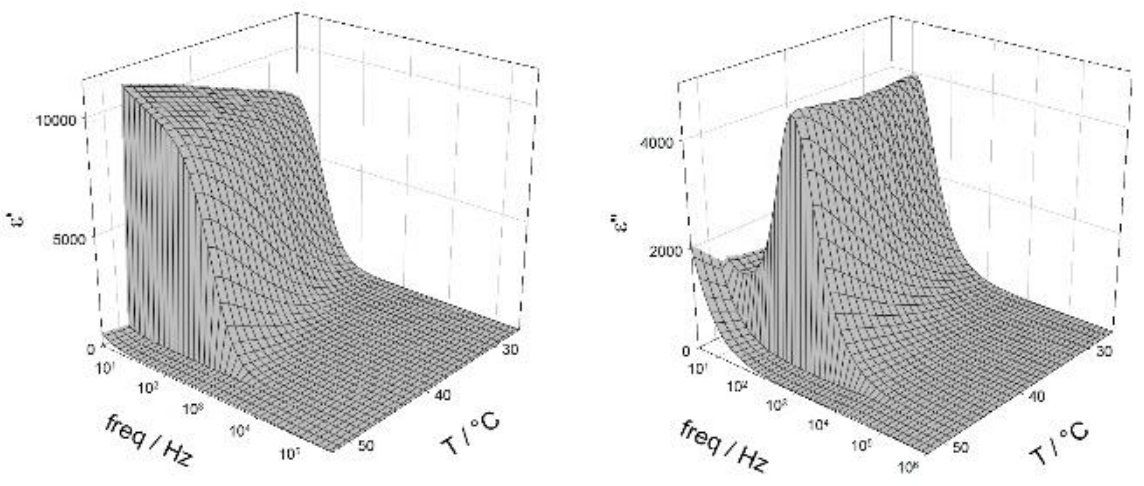

**Figure S10.** Real and imaginary part of dielectric permittivity measured vs. temperature and frequency, in 5-μm-thick cells for **II-6**.

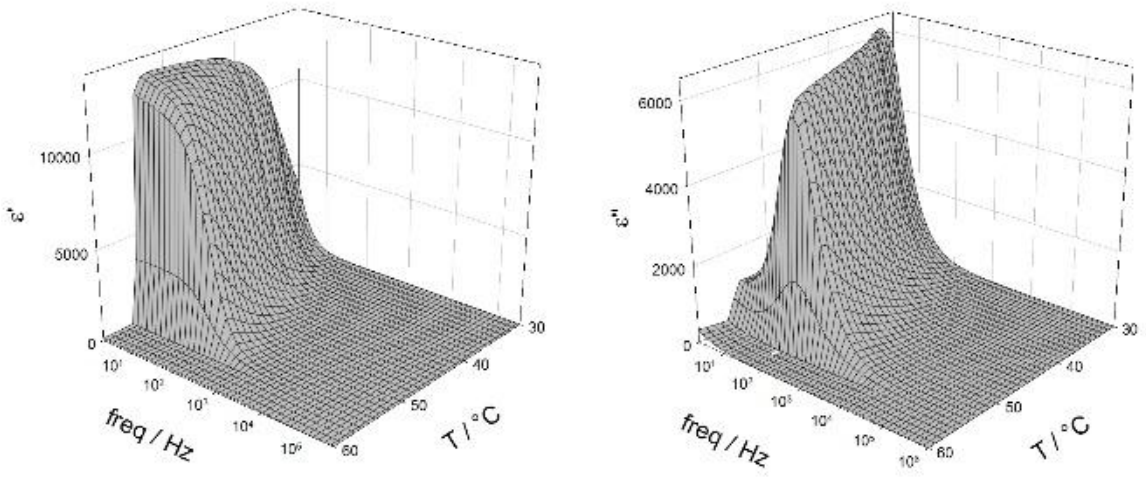

**Figure S11.** Real and imaginary part of dielectric permittivity measured vs. temperature and frequency, in 5-μm-thick cells for **II-4***.

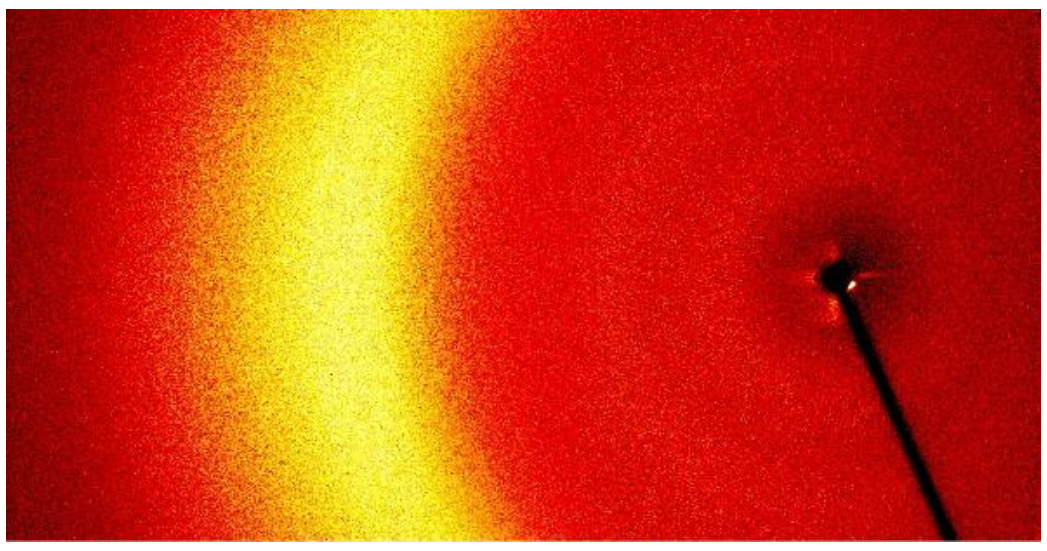

**Figure S12.** 2D X-ray diffraction pattern for compound **I-4*** in N*$_F$ phase at room temperature. The diffused signals, evidencing fully fluidic nature of the phase correspond to periodicities 18.2; 9.5 and 4.5 A.

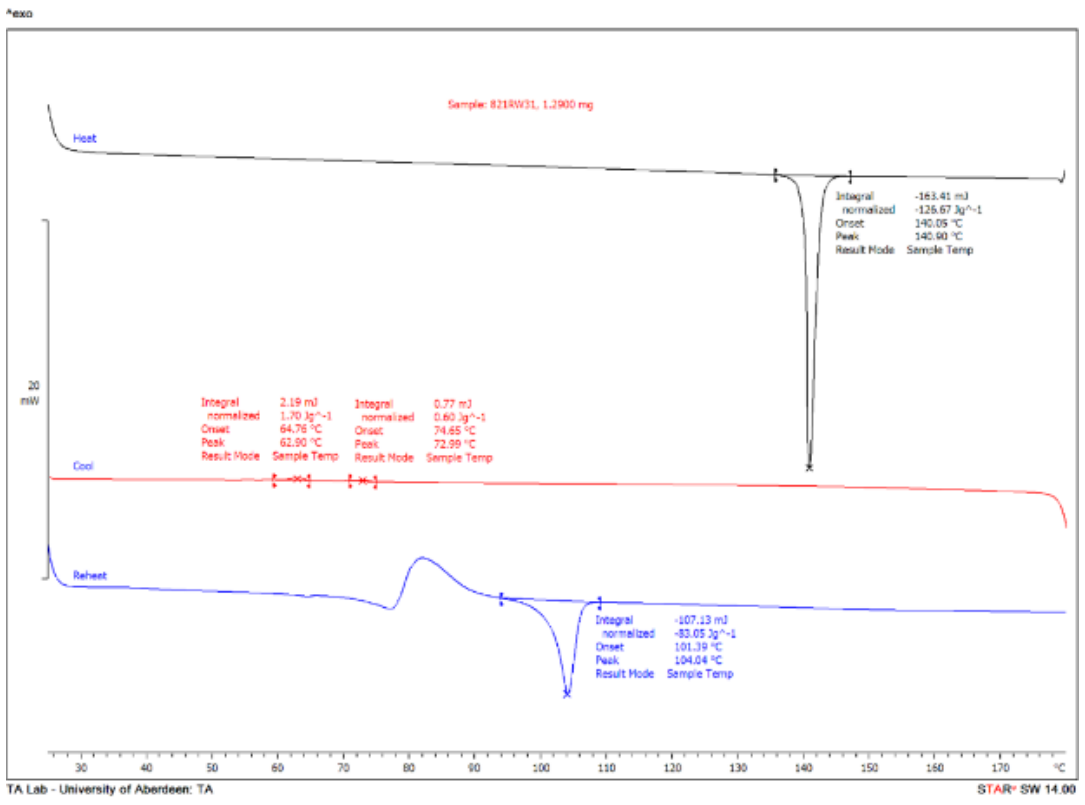

**Figure S13. DSC curve for compound I-4**

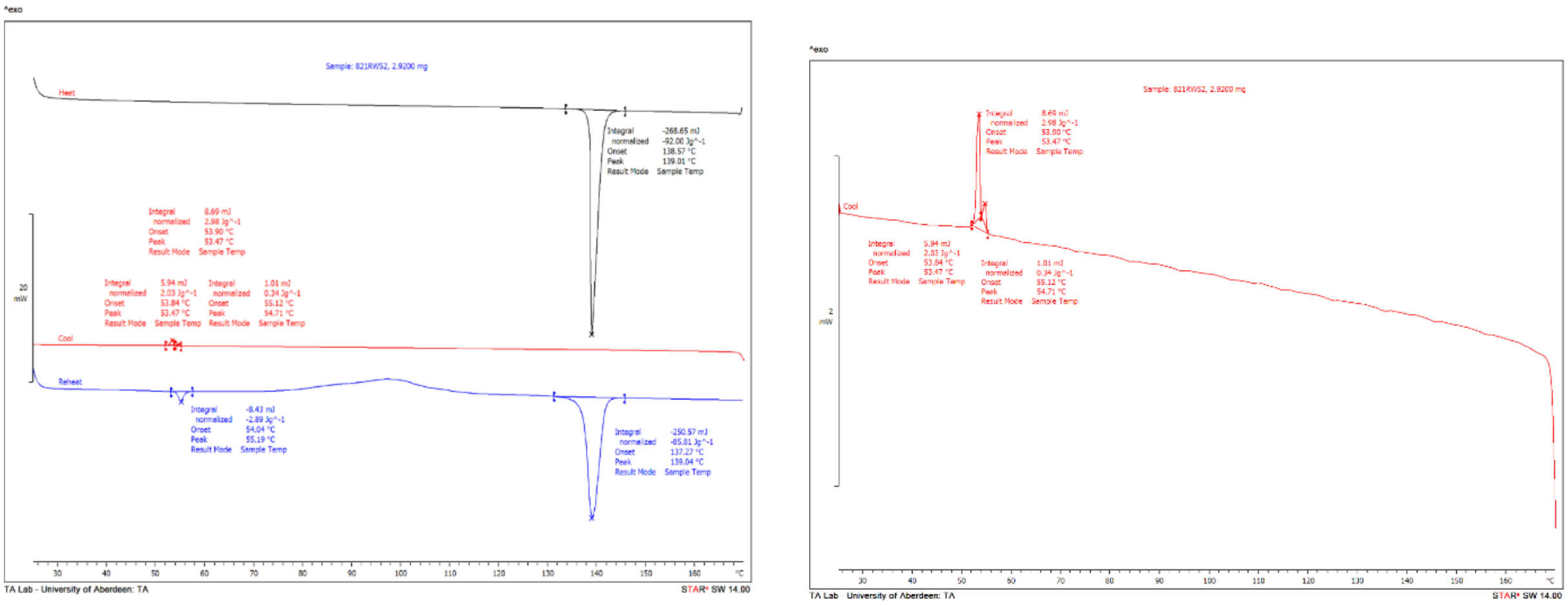

**Figure S14. DSC curve for compound I-4* (left - magnified part of the thermogram showing Iso-N-N*$_F$ transitions)**

**<u>5. References</u>:**